\documentclass[a4paper,12pt]{article}
\usepackage{jheppub}
\usepackage[export]{adjustbox}
\usepackage{color}
\usepackage{tikz}
\usepackage{graphicx}
\usepackage{dcolumn}
\usepackage{bm}
\usepackage{hyperref}
\usepackage{float}
\usepackage[caption=false]{subfig}
\usepackage{hyperref}
\hypersetup{
    colorlinks=true,
    linktoc=page,    
    linkcolor=blue,
    urlcolor=magenta
}
\usepackage{orcidlink}
\newcommand{\be}{\begin{equation}}
\newcommand{\ee}{\end{equation}}
\newcommand{\bea}{\begin{eqnarray}}
\newcommand{\eea}{\end{eqnarray}}
\newcommand{\dd}{\text{d}}
\newcommand{\bra}[1]{\left<{#1}\right|}
\newcommand{\ket}[1]{\left|{#1}\right>}
\newcommand{\ave}[1]{\left<{#1}\right>}
\newcommand{\rh}{r_\text{h}}
\newcommand{\h}{\text{H}}
\newcommand{\N}{\text{N}}
\newcommand{\gen}{\text{gen}}
\newcommand{\tr}{\text{tr}}

\begin{document}
\title{Thermodynamics of the Kerr-AdS black hole from an ensemble-averaged theory}
\author{Peng Cheng\orcidlink{0000-0001-5704-7668}$^{1}$,}
\emailAdd{pchengcn@tju.edu.cn}
\author{Jindong Pan\orcidlink{0009-0003-7232-8152}$^{2}$,}
\emailAdd{jindong.pan@student.uva.nl}
\author{Haichen Xu\orcidlink{0009-0006-0156-9583}$^{2}$}
\emailAdd{haichen.xu@student.uva.nl}
\author{and Si-Jiang Yang\orcidlink{0000-0002-8179-9365}$^{3}$}
\emailAdd{yangsj@lzu.edu.cn (corresponding author)}

\affiliation{1) Center for Joint Quantum Studies and Department of Physics, School of Science,
Tianjin University, Tianjin 300350, China\\
2) Institute for Theoretical Physics, University of Amsterdam, 1090 GL Amsterdam, Netherlands\\
3) Lanzhou Center for Theoretical Physics, Key Laboratory of Theoretical Physics of Gansu Province, and Key Laboratory of Quantum Theory and Applications of MoE, Lanzhou University, Lanzhou, Gansu 730000, China
}
\abstract{
Exploring the universal structure of the gravitational path integral beyond semi-classical saddles and uncovering a compelling statistical interpretation of black hole thermodynamics have long been significant challenges.
We investigate the statistical interpretation of the Kerr-AdS black hole thermodynamics through an ensemble-averaged theory.
By extending the phase space to include all possible states with conical singularities in their Euclidean counterparts, we derive the probability distribution of different states inherited from the Euclidean gravitational path integral.
Moreover, we can define a density matrix of all states in the phase space.
By ensemble-averaging over all states, we show that the black hole phase transition naturally arises in the semi-classical limit.
Away from the semi-classical regime, the ensemble-averaged theory exhibits a notable deviation from the conventional phase transition.
Expanding around the classical saddles yields the subleading-order correction to the Gibbs free energy, which is half of the Hawking temperature.
We demonstrate that the half Hawking temperature correction is a universal feature inherent to black holes in asymptotically AdS spacetime.
With the subleading-order correction to Gibbs free energy, we also suggest that the whole black hole thermodynamic should be corrected accordingly.}

\maketitle
\flushbottom


\section{Introduction}\label{sec: intro}
Black hole thermodynamics has become one of the most intriguing research topics since its birth in the 1970s. Black holes not only are strong gravitational systems but also exhibit thermodynamic properties. They possess entropy and temperature~\cite{Bekenstein:1973ur,Bardeen:1973gs,Bekenstein:1975tw}. These properties may emerge from underlying microscopic structures~\cite{Strominger:1996sh,Maldacena:1996gb,Wei:2015iwa}. Recently, the thermodynamics of black holes in AdS spacetime has garnered considerable attention, not only for its rich phase structure but also for its implications within the framework of the Anti-de Sitter/conformal field theory (AdS/CFT) correspondence.

The thermodynamics of black holes in AdS spacetime unveils a rich tapestry of intriguing phase structures and has experienced remarkable advancements over the past four decades. The Hawking-Page transition is a phase transition between thermal AdS spacetime and black holes~\cite{Hawking:1982dh}. The first-order phase transition for a Schwarzschild-AdS black hole immersed in a thermal bath of radiation can be interpreted as the confinement/deconfinement phase transition in the dual quark-gluon plasma~\cite{Witten:1998qj,Witten:1998zw}. The phase structure for the charged AdS black hole resembles the phase behavior of a Van der Waals fluid~\cite{Chamblin:1999hg,Chamblin:1999tk}. When the negative cosmological constant is interpreted as the thermodynamic pressure~\cite{Kastor:2009wy}, the phase behavior for the charged AdS black hole mirrors that of a Van der Waals fluid, and both share the same critical exponents~\cite{Kubiznak:2012wp}. The thermodynamics of many other AdS black holes resembles the phase behavior of everyday thermodynamic systems, exhibiting features such as triple points~\cite{Altamirano:2013uqa,Altamirano:2013ane} and reentrant phase transition~\cite{Gunasekaran:2012dq,Frassino:2014pha}, among others. The interpretation of a dynamic negative cosmological constant has achieved great success from different perspectives, including the extended Iyer-Wald formalism~\cite{Xiao:2023lap}, the coupling constant of gauge field~\cite{Hajian:2023bhq}, and higher dimensional origins~\cite{Frassino:2022zaz}. For a different interpretation of the cosmology constant, see~\cite{Karch:2015rpa,Mancilla:2024spp}.

Despite great achievements in black hole thermodynamics, the statistical interpretation of the black hole phase transition still remains elusive.
The challenge of quantum gravity remains an unsolved enigma, but the association of a thermal partition function to a black hole system allows the weight of state to be derived from an Euclidean path integral over spacetime geometries, as proposed by Gibbons and Hawking~\cite{Gibbons:1976ue}.
This may be the sally port for a better understanding of the phase space beyond classical geometries.

From a semiclassical perspective, the primary contribution to the path integral arises from the extremums of the action, commonly referred to as the saddle point approximation. Given the exponential dependence on the action, the partition function is predominantly determined by the point at which the Euclidean action attains its global minimum.
For example, the Hawking-page phase transition in AdS spacetime~\cite{Hawking:1982dh} corresponds to a shift of the global minimum of the Euclidean action, where the Euclidean action is the on-shell action.
The black hole thermodynamics only captures the contributions from the saddle points, which are on-shell geometries.
Finding more states away from the saddle points is one of the critical steps toward Euclidean quantum gravity.
Recently, the concept of the free energy landscape was proposed as a new approach to understanding black hole phase transitions~\cite{Li:2020khm,Li:2020nsy}, by adding off-shell geometries. In particular, on-shell black hole solutions correspond to extreme points of the generalized free energy. It was further shown that the generalized free energy can be derived via the gravitational path integral by considering geometries with conical singularities~\cite{Li:2022oup}.

The free energy landscape provides a visible way of understanding the black hole phase transition by comparing the depth of the generalized free energy~\cite{Yang:2021ljn}.
It was a qualitative description until the introduction of the ensemble average in~\cite{Cheng:2024hxh}.
Inspired by the generalized free energy, we can define the density matrix related to Euclidean geometries with or without conical singularities. The probability distribution of the states can be directly calculated through the gravitational path integral.
This distribution facilitates the definition of ensemble average over all the above-mentioned states in the phase space.
It was shown in~\cite{Cheng:2024hxh} that, in the small $G_{\N}$ (semi-classical) limit, the ensemble-averaged theory reduces to classical thermodynamics for Schwarzschild-AdS and Reissner-Nordstr\"om-AdS (RN-AdS) black holes.
Beyond the semi-classical limit, the quantum effects associated with off-shell geometries exhibiting conical singularities become significant.
Furthermore, it was shown that the leading-order $G_{\N}$ contribution of the averaged theory is the black hole thermodynamics, while the subleading-order contribution can be assessed through both numerical and analytical methods, yielding results equivalent to half of the Hawking temperature. Notably, the subleading-order corrections for both black holes are identical.

It is a significant development to quantitatively derive the black hole thermodynamics from an ensemble-averaged theory with a well-defined density matrix. Moreover, the subleading-order correction captures the effects of the off-shell geometries.
As the most common model of the black hole in nature, we are curious about the ensemble-averaged theory of the Kerr-AdS black hole and the statistical interpretation of its thermodynamics.
In this paper, we extend the investigation to the Kerr-AdS case and study whether the subleading-order correction from the previous study is a generic feature.
As the setup, we put the Kerr-AdS black hole in a thermal cavity with ensemble temperature $T_{\text{E}}$.
We should allow all possible bulk geometries as long as they are described by the same Lorentzian physics. With the fixed ensemble temperature and Lorentzian physics, we can have black holes of different sizes in bulk.
The situation with the largest probability is the case when the ensemble temperature $T_{\text{E}}$ equals the Hawking temperature $T_{\text{H}}$ calculated from the surface gravity $\kappa$.
For other black hole sizes, the corresponding Hawking temperature $T_{\text{H}}$ does not equal the ensemble temperature $T_{\text{E}}$, resulting in conical singularities in their Euclidean geometries, and the weights of those states can be evaluated.
Ensemble averaging over those states in the phase space enables us to naturally derive the black hole phase transition and to reveal additional physics beyond the classical regime.

The structure of the paper is as follows.
In Sec.~\ref{sec:landscape}, we review the free energy landscape in the context of Kerr-AdS black holes. In Sec.~\ref{sec: average}, we define the ensemble-averaged free energy via the gravitational path integral and give a statistical interpretation of black hole thermodynamics.
We numerically compute the ensemble-averaged free energy for different values of $G_{\N}$ and conclude that black hole thermodynamics is a small $G_{\N}$ effect.
In Sec.~\ref{sec:subleading}, we expand the ensemble-averaged free energy in different orders of $G_{\N}$ and find the subleading corrections via numerical and analytical methods. Using these two approaches, we derive the result that the first-order quantum correction to the free energy is half of the Hawking temperature.
We also prove the claim that the first-order quantum correction is universal for all asymptotic AdS black holes and discuss the concept of quantum-corrected black hole thermodynamics.
The last section is devoted to the conclusion and discussion.

\section{Thermodynamics and free energy landscape of the Kerr-AdS black hole}\label{sec:landscape}

The Kerr-AdS black hole is a vacuum solution of Einstein's field equation with a negative cosmological constant, describing a rotating black hole in AdS spacetime. In this section, we review the thermodynamics and free energy landscape of the Kerr-AdS black hole.

\subsection{Thermodynamics of Kerr-AdS black hole}

The metric for the four-dimensional Kerr-AdS black hole takes the form
\begin{align}
    d s^2=-\frac{\Delta}{\rho^2}\left(\dd t-\frac{a\sin^2\theta}\Xi \dd \phi\right)^2+\frac{\rho^2}\Delta \dd r^2+\frac{\rho^2}{\Sigma} \dd\theta^2+\frac{\Sigma\sin^2\theta}{\rho^2}\left(a\dd t-\frac{r^2+a^2}\Xi \dd\phi\right)^2,\label{metric}
\end{align}
with the metric functions
\begin{equation}
    \begin{split}
     \Delta &=(r^2+a^2)\left(1+\frac{r^2}{L^2}\right)-2G_{\N} mr,  \qquad  \Xi=1-\frac{a^2}{L^2}, \\
     \rho^2 &=r^2+a^2\cos^2\theta, \qquad \qquad \qquad \quad \qquad \Sigma=1-\frac{a^2}{L^2}\cos^2\theta,
    \end{split}
\end{equation}
where $L$ is the AdS radius,
$m$ and $a$ are the mass and angular momentum parameters, respectively. We preserve Newton's constant $G_{\N}$ for the convenience of later discussion of the quantum correction for black hole thermodynamics. The Kerr-AdS spacetime is a steady and axisymmetrical spacetime with ADM mass and angular momentum~\cite{Gibbons:2004ai,Henneaux:1985tv}:
\be
M=\frac{m}{\Xi^{2}},\quad \quad J=\frac{ma}{\Xi^{2}}.
\ee
The event horizon radius denoted as $r_\h$ is the largest solution to equation $\Delta=0$.
Writing physical quantities in terms of the horizon radius $r_\h$, the mass and angular momentum of the Kerr-AdS black hole can be expressed as:
\begin{align}
    M &= \frac{r_\h}{2G_{\N} \Xi^2}\left(1+\frac{a^2}{r_\h^2}\right)\left(1+\frac{r_\h^2}{L^2}\right), \\
    J &= \frac{a r_\h}{2G_{\N} \Xi^2}\left(1+\frac{a^2}{r_\h^2}\right)\left(1+\frac{r_\h^2}{L^2}\right)\label{eq:J}.
\end{align}
The Hawking temperature and entropy of the black hole are
\begin{align}
    T_\h &=\frac{r_\h}{4\pi\left(r_\h^2+a^2\right)} \left(1+\frac{a^2}{L^2}+\frac{3r_\h^2}{L^2}-\frac{a^2}{r_\h^2}\right),\label{THH}\\
S &= \frac{\pi(r_\h^2+a^2)}{G_{\N} \Xi }.
\end{align}
The Kerr-AdS spacetime describes a rotating black hole with angular velocity
\begin{equation}
    \Omega_{\h}=\frac{a}{L^2}\frac{r^2_\h+L^2}{r^2_\h+a^2}.\label{angularV}
\end{equation}

In the extended phase space thermodynamics, the cosmological constant $\Lambda$ is interpreted as the thermodynamic pressure, and the mass is interpreted as the enthalpy instead of internal energy~\cite{Kastor:2009wy}. The pressure of the black hole is
\begin{equation}
P=-\frac{\Lambda}{8\pi G_{\N}}= \frac{3}{8\pi G_{\N}}\frac{1}{L^2}. \label{pressure}
\end{equation}
Here we preserved Newton's constant $G_{\N}$ in the thermodynamic pressure $P$.
The thermodynamic volume of the black hole is
\begin{equation}
 V=\frac{4\pi}{3}\frac{r_\h(r_\h^2+a^2)}{\Xi}\left(1+\frac{a^2}{2r_\h^2}\frac{1+\frac{r_\h^2}{L^2}}{\Xi}\right).
\end{equation}
The thermodynamic volume can be defined as a surface integral of the Killing potential~\cite{Kastor:2009wy} or defined as the Killing volume~\cite{Jacobson:2018ahi,Ahmed:2023dnh}
\begin{equation}
    V=\int_{\Sigma_{\text{bh}}}\mid\mid \xi \mid\mid d V-\int_{\Sigma_{\text{AdS}}}\mid\mid \xi \mid\mid d V,
\end{equation}
where $\mid\mid \xi \mid\mid=\sqrt{-\xi \cdot \xi}$ is the norm of the event horizon generating Killing vector $\xi=\partial_t+\Omega_{\text{H}}\partial_{\phi}$.

From the above thermodynamic variables, we can get the following first law of black hole thermodynamics and the Smarr relation:
\begin{align}
    dM&=T_{\h}dS+\Omega_{\h}dJ+VdP; \label{1stlaw} \\
    M&=2T_{\h}S+2\Omega_{\h} J-2PV. \label{SmarrR}
\end{align}
With the first law of thermodynamics for the Kerr-AdS black hole established, we can investigate its thermodynamic phase behavior, which is often described by the Gibbs free energy. The corresponding Gibbs free energy of the black hole is
\be
\begin{split}
F_{\text{Gibbs}}&=M-T_\h S \\
&=\frac{1}{4G_{\N} \Xi r_\h}\left[a^2\Big(1-\frac{r_\h^2}{L^2}\Big)+\frac{2(a^2+r_\h^2)(L^2-r_\h^2)}{L^2-a^2}-r_\h^2\Big(1+3\frac{r_\h^2}{L^2}\Big)\right]\,.\label{FGibbs}
\end{split}
\ee

\begin{figure}[h]
    \centering
    \includegraphics[width=0.6\textwidth]{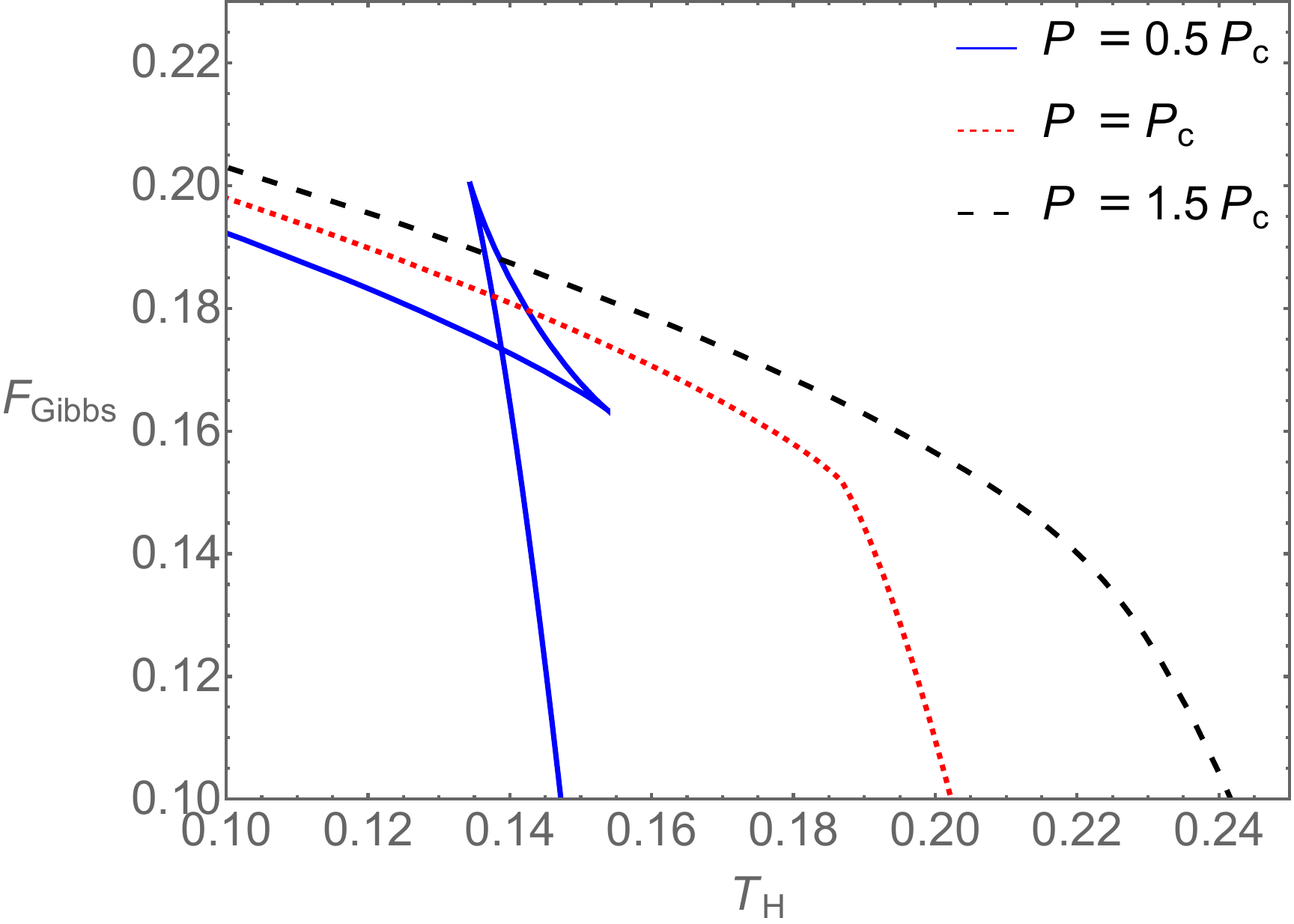} 
    \caption{The Gibbs free energy for Kerr-AdS black hole with fixed $J=0.05/G_{\N}$ and $G_{\N}=1$ for various values of the pressure $P$. The red line corresponds to the pressure $P=0.5 P_c$, and the Gibbs free energy $F_{\text{Gibbs}}$ shows the characteristic swallowtail behavior which indicates a phase transition from a small black hole to a large black hole. When the pressure $P=P_c$, there exists a continuous phase transition. There is no phase transition for pressure $P=1.5 P_c$.}
    \label{fig:example}
\end{figure}

A phase transition of a black hole is described by the Gibbs free energy $F_{\text{Gibbs}}$. If the first derivative of the Gibbs free energy with respect to temperature is discontinuous, the transition is classified as first-order. From Fig.~\ref{fig:example} for the Gibbs free energy $F_{\text{Gibbs}}$ vs black hole temperature $T_{\h}$,
there are characteristic swallowtail behaviors for pressures below the critical pressure $P_c$, indicating a first-order phase transition between small and large black holes. The swallowtail behavior disappears for pressure larger than the critical pressure $P_c$. The critical point of the phase transition can be derived from the $T_{\h}-S$ oscillatory curve. It satisfies
\begin{equation}
    \begin{split}
       \left( \frac{\partial T_{\h}}{\partial S}\right)_{J,P}=0, \qquad \left(\frac{\partial^2 T_{\h}}{\partial S^2}\right)_{J,P}=0,
    \end{split}
\end{equation}
which is~\cite{Wei:2015ana}
\begin{equation}
P_c =\frac{\alpha _1}{G_{\text{N}}^2J},\quad \quad T_{c}=\frac{\alpha _2}{\sqrt{G_{\text{N}} J} }\,,
\end{equation}
with $\alpha_1\approx 0.002857$ and $\alpha_2\approx 0.0041749$.

The phase transition of the Kerr-AdS black hole shares many similarities with the liquid-gas phase transition observed in daily life, and its critical exponents are identical to those of a Van der Waals fluid\cite{Gunasekaran:2012dq, Altamirano:2014tva, Wei:2015ana}.

\subsection{Free energy landscape of the Kerr-AdS black hole}
\label{sec22}

The Gibbs free energy characterizes the preferred phase of a thermodynamic system. To gain insight into the physical processes underlying phase transitions and thermodynamic stability, we examine the free energy landscape.
In this section, we cover some important developments related to the free energy landscape.
When geometries beyond the classical AdS or black hole solutions are considered in the Euclidean path integral formulation of gravity, the corresponding free energy derived from the Euclidean action is referred to as the generalized free energy.
This generalized free energy~\cite{York:1986it} proposed by York, is a key concept for describing the quantum fluctuation of the black hole, which can result in the phase transition of the black hole.

In recent works~\cite{Li:2021vdp,Li:2022oup,Li:2022ylz,Li:2022yti,Wei:2021bwy,Li:2023men,Yang:2023xzv}, it was shown that
Euclidean geometries with conical singularities play an important role in understanding the black hole phase transition and thermodynamic stability.
The generalized free energy, calculated by incorporating conical singularities, offers valuable insights into black hole phase transitions.
First, we review the derivation of the generalized free energy from the Euclidean path integral and then explain how the free energy landscape provides insights into phase transitions.

\subsubsection{Euclidean path integral}

\begin{figure}[!htb]
\begin{center}
  \subfloat[The Euclidean picture]{\includegraphics[width=0.5 \textwidth]{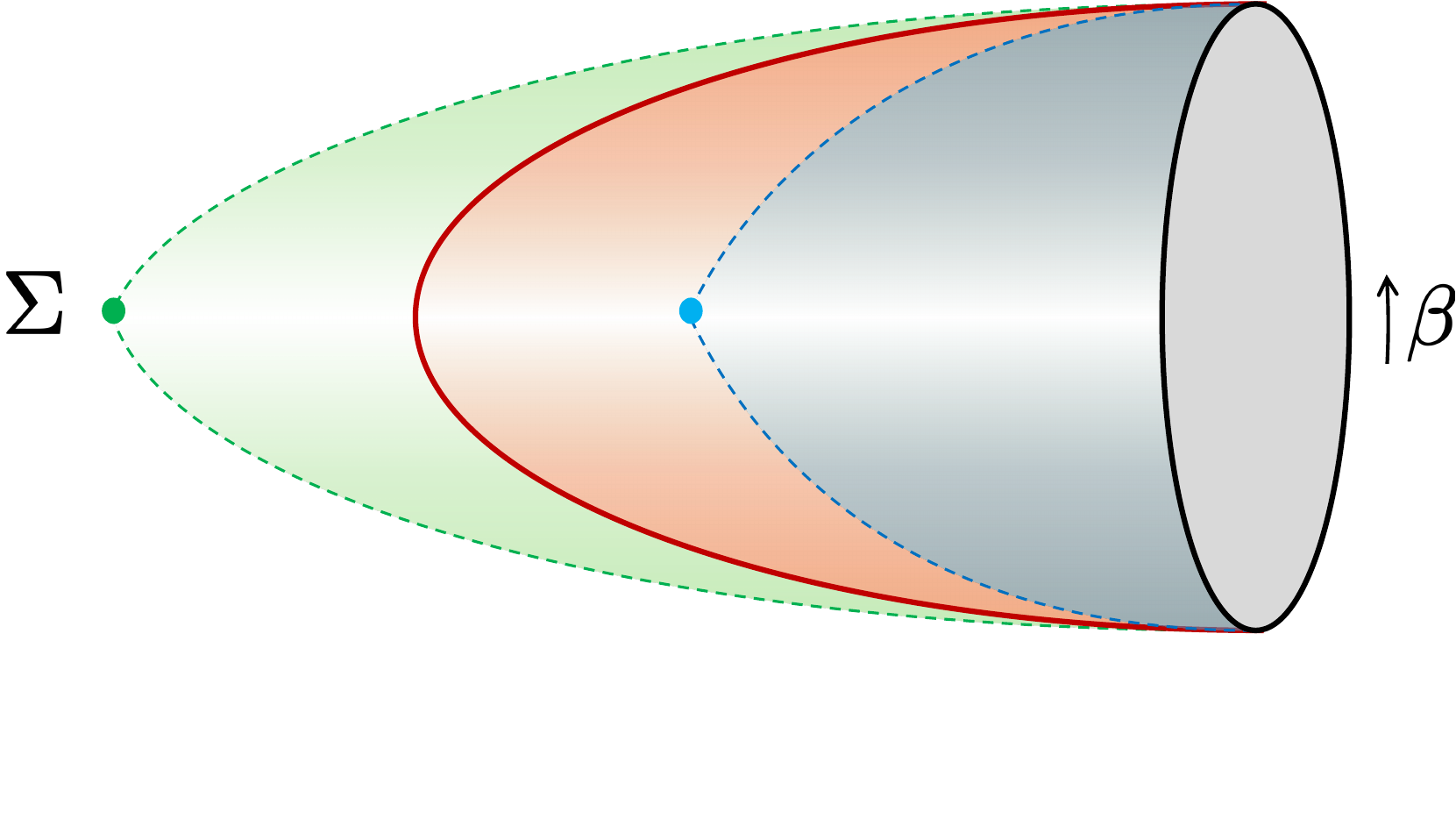}\label{cigara}}~~~~~~~~~
  \subfloat[The Lorentzian picture]{\includegraphics[width=0.35 \textwidth]{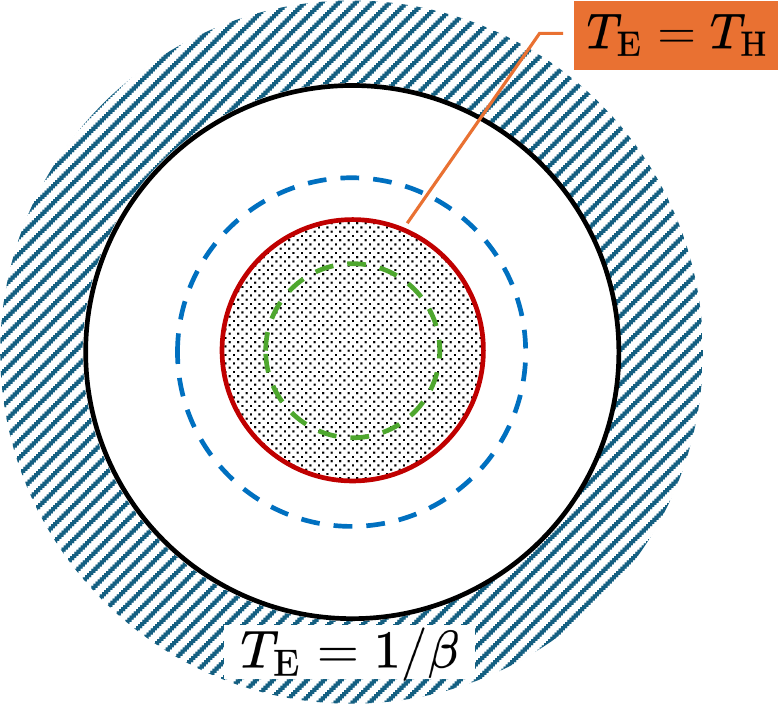}\label{cigarb}}
    \caption{(a) The Euclidean geometries included in the phase space. The orange cigar geometry with a smooth tip corresponds to a black hole in the Lorentzian signature, while the others are the ones with conical singularities.
    (b) The Lorentzian setup. Putting a black hole in a cavity with boundary ensemble temperature $T_{\text{E}}=1/\beta$, the smooth geometry corresponds to the black hole case with $T_{\text{E}}=T_\h$. When $T_{\text{E}} \neq T_\h$, there would be conical singularities on its Euclidean geometry.}
    \label{cigars}
 \end{center}
\end{figure}

Imagine putting a black hole in a thermal cavity with wall temperature $T_{\text{E}}=1/\beta$ as shown in Fig. \ref{cigarb}.
We should always suppose that the bulk equilibrium is guaranteed.
For any horizon radius, there is a Hawking temperature $T_\h$ calculated from the surface gravity $\kappa/2\pi$, which is expressed as \eqref{THH}.
When the ensemble temperature $T_{\text{E}}$ equals the Hawking temperature of the black hole, there is no conical singularity.
The corresponding Euclidean geometry is the orange cigar in Fig. \ref{cigara}.
There can be off-shell geometries corresponding to black holes with different sizes.
For the case that the Hawking temperature calculated from the horizon radius does not match $T_\text{E}$, there will be a conical singularity at the tip of the Euclidean geometry as can be seen from Fig. \ref{cigara}.
The blue and green geometries with singularity correspond to geometries with different horizon temperatures.

The action of the Euclidean geometries with conical singularities at their tips can be directly evaluated~\cite{Fursaev:1995ef,Solodukhin:1994yz,Mann:1996bi,Solodukhin:2011gn,Li:2022oup}.
The Euclidean Einstein-Hilbert action is~\cite{Gibbons:1976ue}
\begin{equation}
    \bar{I}_{\text{E}}=-\frac{1}{16 \pi G_{\N}}\int_{\mathcal{M}}d^4x\  \sqrt{g}   \left(R+\frac{6}{L^2}\right) -\frac{1}{8 \pi G_{\N}}  \int_{\partial \mathcal{M}}d^3 x\ \sqrt{h} K 
    ,\label{EucAction}
\end{equation}
where $R$ is the Ricci scalar curvature,
and $K$ is the extrinsic curvature on the AdS boundary.
We are considering a black hole away from its classical saddle point, i.e., its horizon temperature differs from the Hawking temperature $T_\h $. In this case, the geometry has a conical singularity $\Sigma$ at the horizon~\cite{Mann:1996bi}. It can be shown that the Euclidean action of the geometry with the conical singularity can be written as~\cite{Solodukhin:2011gn}
\begin{equation}
   \int_{\mathcal{M}} d^4 x \sqrt{g} (R+\frac{6}{L^2}) =4\pi (1-\frac{\beta }{\beta _{\h} }) \int _\mathcal{H} d^3 x \sqrt{h}  +\int _{M/\Sigma } d^4 x \sqrt{g} (R+\frac{6}{L^2})\,,
\end{equation}
where $\mathcal{H}$ is the horizon, $h_{ij}$ is the induced metric on the black hole horizon, and $M/\Sigma$ is the regularized manifold. 
The action can be further evaluated as
\begin{equation}
  \bar{I}_{\text{E}} = -\frac{1}{4 G_{\N}}\left( 1-\frac{\beta}{\beta_{\h}} \right)A + \bar{I}_{M/\Sigma}\,,\label{IEE}
\end{equation}
where $A=4 \pi\left(\rh^2+a^2\right)/\Xi$ is the area of the event horizon, and $\bar{I}_{M/\Sigma}$ is the regular part of the action with the same expression as Eq.~\eqref{EucAction}.
The regular part is divergent because the volume of space is infinite. In order to regularize this divergence, we introduce a ``counterterm'', which is the action of the pure AdS spacetime~\cite{Hawking:1982dh}.
In the radial direction, we integrate up to $r_0$, and finally take $r_0 \rightarrow \infty$. To do this, we need to match the boundary of Kerr-AdS spacetime and that of pure AdS spacetime. We set the length of the Euclidean time circle the same for both cases,
indicating that the period of the Euclidean time for both pure AdS spacetime and Kerr-AdS spacetime is related as follows
\begin{equation}\label{eq: beta vs beta0}
\beta_0=\beta\left[1-\frac{m L^2}{r_0^3}+\mathcal{O}\left(\frac{1}{r_0^5}\right)\right].
\end{equation}
The regularized action can be calculated as
\begin{equation}
         \bar{I}_{M/\Sigma} 
    = \frac{3}{8 \pi G_{\N} L^2}\left(\beta \int_{\rh}^{r_0} \sqrt{g} d r d \theta d \phi-\beta_0 \int_0^{r_0} \sqrt{g_0} d r d \theta d \phi\right),
\end{equation}
where $g$ is the determinant of the metric of the Kerr-AdS spacetime, and $g_0$ is the determinant of the metric of the pure AdS spacetime.
Further calculation shows that
\begin{equation}
  \bar{I}_{M/\Sigma} = \frac{\beta}{2G_{\N} \Xi}\left[ m\left(1+\frac{a^2}{r_0^2}\right) - \frac{\rh(\rh^2+a^2)}{L^2} \right]\,,
\end{equation}
where we have used Eq.~(\ref{eq: beta vs beta0}). Taking $r_0 \rightarrow \infty$, we obtain
\begin{equation}
\bar{I}_{M/\Sigma}=\frac{\beta}{2 G_{\N} \Xi}\left[m-\frac{\rh\left(\rh^2+a^2\right)}{L^2}\right]\,.\label{IMS}
\end{equation}
Inserting Eq.~\eqref{IMS} in Eq.~\eqref{IEE}, the Euclidean action of the geometry with a conical singularity is
\begin{equation}
\bar{I}_{\text{E}}  =\frac{\beta \rh}{2 G_{\N} \Xi}\left(1+\frac{\rh^2}{L^2}\right)-\frac{\pi\left(\rh^2+a^2\right)}{G_{\N} \Xi}.
\end{equation}
Transferring to a new ensemble with fixed $J$, we have
\be
I_{\text{E}}=\bar{I}_{\text{E}}+\beta\Omega_{\h} J=\frac{\beta \rh}{2 G_{\N} \Xi^2}\left(1+\frac{a^2}{\rh^2}\right)\left(1+\frac{\rh^2}{L^2}\right)-\frac{\pi\left(\rh^2+a^2\right)}{G_{\N} \Xi}\label{IE}\,.
\ee
Therefore, the generalized free energy for the Kerr-AdS black hole can finally be written as
\begin{equation}
F_{\gen}=\frac{\rh}{2 G_{\N} \Xi^2}\left(1+\frac{a^2}{\rh^2}\right)\left(1+\frac{\rh^2}{L^2}\right)-\frac{\pi T_\text{E}\left(\rh^2+a^2\right)}{G_{\N} \Xi}.\label{genFEnergy}
\end{equation}

\begin{figure}[H]
    \centering
    \subfloat[]{\includegraphics[scale=0.38]{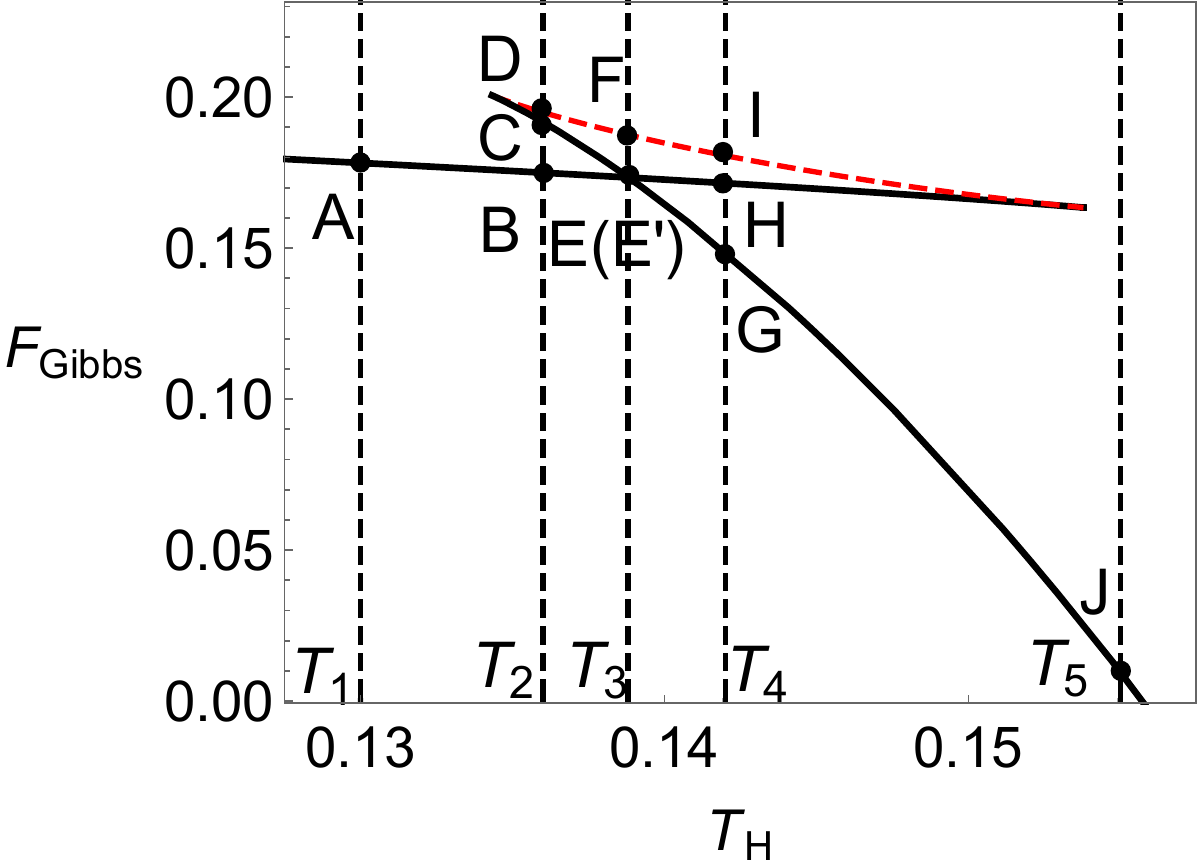}\label{Fig:Gibbsa}}~~~~
    \subfloat[$T_{\text{E}}=T_1=0.130$]{\includegraphics[scale=0.43]{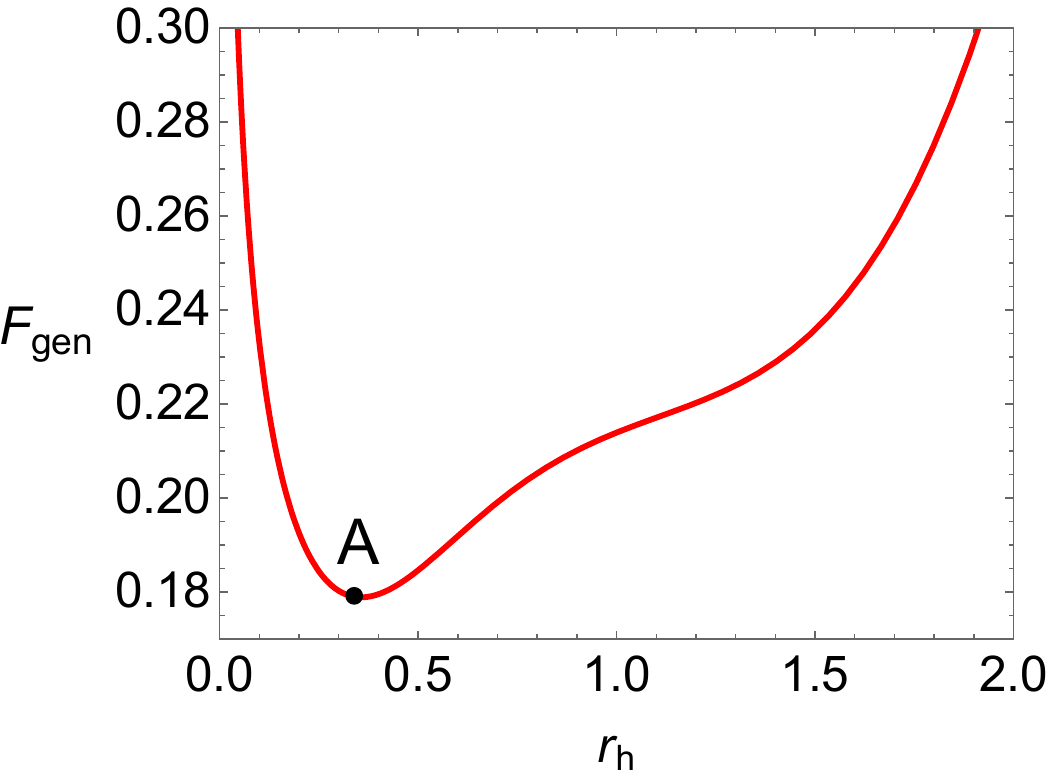}\label{Fig:fGibbsb}}
\\
    \subfloat[$T_{\text{E}}=T_2=0.136$]{\includegraphics[scale=0.425]{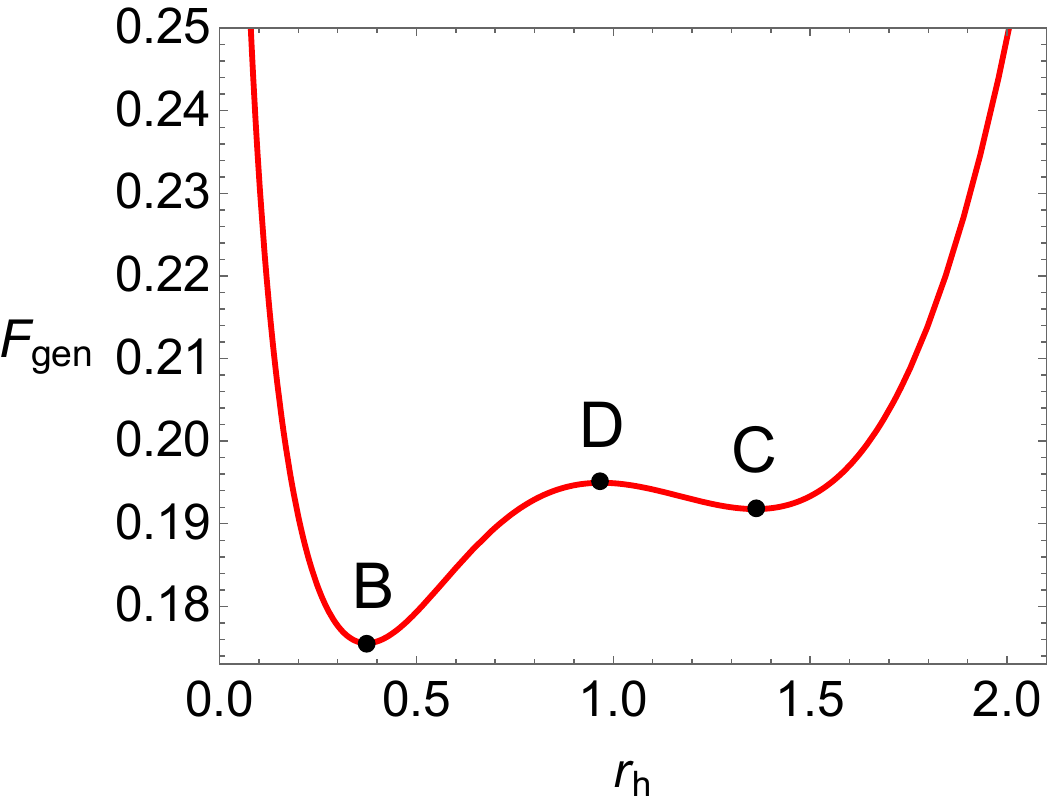}\label{Fig:fGibbsc}}~~~~
    \subfloat[$T_{\text{E}}=T_3=0.139$]{\includegraphics[scale=0.43]{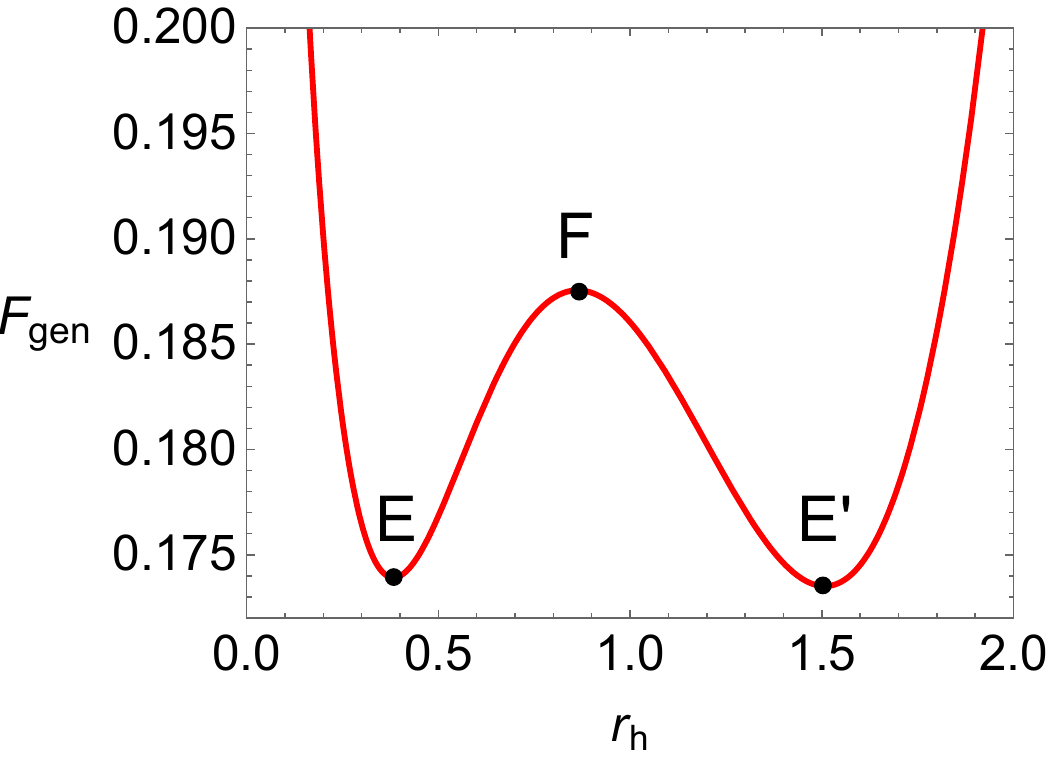}\label{Fig:fGibbsd}}
\\
    \subfloat[$T_{\text{E}}= T_4=0.142$]{\includegraphics[scale=0.42]{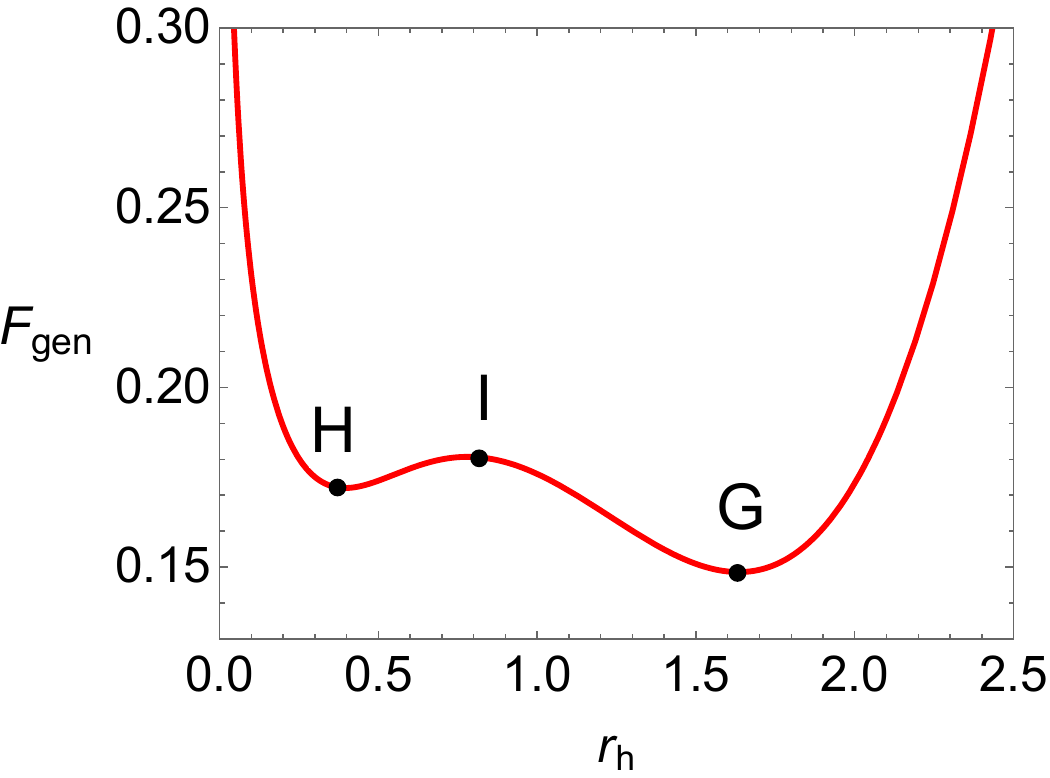}\label{Fig:fGibbse}}~~~~
    \subfloat[$T_{\text{E}}=T_5=0.155$]{\includegraphics[scale=0.42]{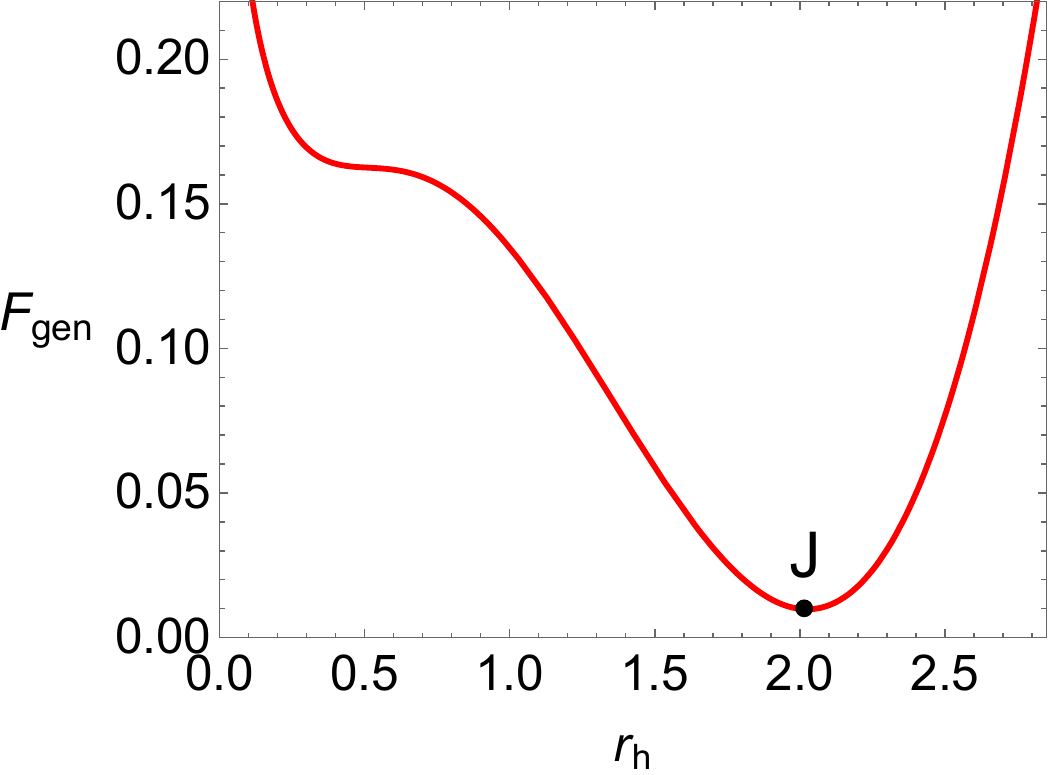}
    \label{Fig:fGibbsf}}
    \caption{The swallow tail behavior of the Gibbs free energy and corresponding generalized free energy, where we set the Newton constant $G_{\N}=1$, angular momentum $J=0.05$ and the thermodynamic pressure $P=0.5P_c$. The swallow tail corresponds to the double well configuration of $F_{\gen}$.}
    \label{Fig:FGibbs}
\end{figure}

Note that the generalized free energy~\eqref{genFEnergy} is generally different from the black hole Gibbs free energy~\eqref{FGibbs}.
However, the generalized free energy~$F_{\text{gen}}$ matches the Gibbs free energy of the black hole when the ensemble temperature $T_{\text{E}}$ equals the black hole temperature $T_{\h}$.
This is the situation when $\rh$ equals the horizon radius $r_\h$ without conical singularity and it corresponds to the extreme points of the off-shell free energy, as shown in Fig.~\ref{Fig:FGibbs}.

\subsubsection{Black hole free energy landscape}

The generalized free energy~\eqref{genFEnergy}
provides a visible picture of the black hole phase transition.
To visualize the behavior of the generalized free energy, we plot the generalized free energy $F_{\gen}$ versus the horizon radius $\rh$. To illustrate its relation to the Gibbs free energy $F_{\text{Gibbs}}$ of the black hole, we also plot the Gibbs free energy $F_{\text{Gibbs}}$ versus the Hawking temperature $T_\h$. As we discussed previously, the Gibbs free energy exhibits a characteristic swallow-tail behavior for thermodynamic pressure $P$ below the critical pressure $P_c$. The five dashed vertical lines in Fig.~\ref{Fig:Gibbsa} correspond to five different temperatures. The black points in the swallowtail correspond to the equilibrium phases of the black hole at different ensemble temperatures. As illustrated in Figs.~\ref{Fig:fGibbsb} to~\ref{Fig:fGibbsf}, for low and high ensemble temperatures, there is only one black hole phase, corresponding to the minimum of the well. At intermediate ensemble temperatures, the generalized free energy exhibits a double well, resulting in three equilibrium black hole phases, which correspond to the three extremal values of the generalized free energy.

As demonstrated in previous work~\cite{Wei:2020rcd,Yang:2021ljn}, the extreme points of the generalized free energy correspond to black holes in equilibrium with the environment. From the definition of the generalized free energy and the first law of black hole thermodynamics, we have
\begin{equation}
    \begin{split}
        \left(\frac{\partial F_{\gen}}{\partial \rh}\right)_{J,P,T_{\text{E}}} =\left[\left( \frac{\partial M}{\partial S}\right)_{J,P,T_{\text{E}}} - T_{\text{E}}\right]\left(\frac{\partial S}{\partial \rh}\right)_{P,J} = \left(T_\h  - T_{\text{E}}\right)\left(\frac{\partial S}{\partial \rh}\right)_{P,J}.
    \end{split}
\end{equation}

The thermodynamic stability of the black hole is related to the second-order derivative of the generalized free energy with respect to the entropy
\begin{equation}
    \begin{split}
     \left.  \left( \frac{\partial^2 F_{\gen}}{\partial \rh^2}\right)_{P,J,T_{\text{E}}}\right|_{T_{\text{E}}=T_{\text{H}}}=\left(\frac{\partial T_{\text{H}}}{\partial S}\right)_{J,P}\left(\frac{\partial S}{\partial \rh}\right)_{P,J}^2=\frac{T_{\text{H}}}{C_{J,P}}\left(\frac{\partial S}{\partial \rh}\right)_{P,J}^2,
    \end{split}
\end{equation}
where $C_{J,P}$ is the heat capacity of the black hole at constant angular momentum $J$ and thermodynamic pressure $P$. At the maximum point of the generalized free energy, the heat capacity is negative, indicating thermodynamic instability. In contrast, at the minimum point of the generalized free energy, the heat capacity is positive, indicating thermodynamic stability, at least locally.




\section{Ensemble average}\label{sec: average}

The free energy landscape gives us a nice picture of how the black hole phase transition occurs when off-shell configurations are included.
Although one can see that the global minimums on the landscape describe the black hole thermodynamics, a quantitative description of the phase transition is not provided.
Moreover, our goal for studying black hole thermodynamics is to find a statistical interpretation in terms of the density matrix. We explore these aspects in this section.

\subsection{A statistical interpretation of the black hole thermodynamics}

The customary approach to the many-body theory interpretation of thermodynamics is through the path integral method.
We review the path integral method for bosonic fields in appendix \ref{boson}, and apply the strategy to black hole thermodynamics in this section.

As discussed in appendix \ref{boson}, for bosonic fields, the density matrix and partition function also possess an Euclidean path integral description mimicking the transition amplitude formula
\bea
\bra{\phi_b}e^{-\beta H}\ket{\phi_a}=\int_{\phi(\textbf{x},0)=\phi_a(\textbf{x})}^{\phi(\textbf{x},\beta)=\phi_b(\textbf{x})}[\mathcal{D} \phi]~ \exp\left[-\int_{0}^{\beta}d \tau \int d^3 x ~\mathcal{L}\right]=\int_{\phi_a}^{\phi_b}[\mathcal{D} \phi]~ e^{-I_{\text{E}}}\,,\label{density}
\eea
and
\bea
Z_{\phi}=\sum_{\phi}\bra{\phi}e^{-\beta H}\ket{\phi}=\int_{\text{periodic}}[\mathcal{D} \phi]~ e^{-I_{\text{E}}}\,,\label{Zphi}
\eea
with Euclidean action $I_{\text{E}}$, and ``periodic'' means that we need to sum over all possible configurations that respect the periodic boundary condition in the $\tau$ direction.
The path integral representation of the density matrix implies that the weight of different states in the canonical ensemble can be calculated via the Euclidean path integral, which provides a calculable way of understanding the statistical physics underlying thermodynamics. The method can also be generalized to gravitational physics~\cite{Gibbons:1976ue} and gives insight into the statistical interpretation of black hole thermodynamics.
In the Euclidean path integral formula of gravitational theory, the partition function can be formally represented as
\be
Z_{g}=\int_{\text{periodic}}[\mathcal{D} g]~ e^{-I_{\text{E}}[g]}\,.\label{Zg}
\ee
In principle, we should include all allowed configurations of manifolds $g_{\mu\nu}$ in the path integral \eqref{Zg}, which is extremely complicated.
There are two common ways to make the situation easier to handle.

The first one is the saddle point approximation. It is well-known that the classical on-shell configurations make dominant contributions in the path integral. Hence, the partition function can be approximated as
\be
Z_g \approx \sum_i~e^{-I_{\text{E}}[g^{(i)}_{\mu\nu}]}\times Z_{\phi}\,
\ee
where $g^{(i)}_{\mu\nu}$ are the classical saddles of the action with the subscript $(i)$ labelling different possible solutions, and $Z_{\phi}$ is just the partition function of bosonic fields \eqref{Zphi} representing bosonic fluctuations around the saddles.
This approach, known as the semi-classical approximation, utilizes field fluctuations within fixed classical geometries to elucidate gravitational physics.
In this approximation, one can separately calculate the action of each saddle and then sum up different contributions. When summed over different saddles, if the Euclidean action of one solution surpasses the others, that solution becomes the dominant contribution due to the exponent in Eq.~\eqref{Zg}.
The transition between different dominant saddles corresponds precisely to the black hole phase transition.

The second way to simplify the situation is by adding constraints in the phase space, which means that only a specific setup is being considered.
This is also common in thermal field theory.
When certain physical setups require conserved quantities, or when gauge-fixing conditions need to be imposed, we often insert delta functionals into the path integral to exclude specific physical or unphysical states in the phase space.
Boundary conditions also serve as constraints in the phase space. When performing functional integral with mode expansion, only the modes that respect the boundary condition should be included in the functional integral.

The free energy landscape reviewed in the previous section reveals an interesting circumstance where the saddle points are included and a consistent extension away from the classical saddle is provided.
The saddle points of the gravitational path integral are exactly the extreme points of the free energy landscape because $\exp({-I_{\text{E}}})=\exp({-\beta F_{\text{gen}}})$.
Moreover, there is a set of physical constraints on the phase space, making the system both rich in essential physics and convenient to handle.
The physical fine-grained states included in the phase space all have a fixed boundary ensemble temperature $T_{\text{E}}=1/\beta$ and share the same Lorenzian physics in the bulk. This implies that the Lorentzian metric \eqref{metric} should not be affected by the constraints.
So, with the given ensemble temperature, any physical quantity associated with the metric \eqref{metric} should be an ensemble-averaged quantity
\be
\ave{A}=\frac{1}{Z}~\tr(\rho A)\,,\label{aveA}
\ee
where the density matrix can be computed from the Euclidean path integral \eqref{density}.
The presence of the factor $1/Z$ is due to the fact that the density matrix is not normalized, and one can define normalized density matrix $\hat{\rho}=\rho/Z$ such that the factor in Eq.~\eqref{aveA} is absorbed in the trace.
Following the same logic, the ensemble-averaged free energy can be defined as
\be
\ave{F}=\frac{1}{Z}~\tr(\rho F_{\gen})\,.\label{aveF}
\ee

Let us look at the statistical interpretation of the density matrix and the gravitational path integral.
Imagine putting a black hole in a thermal bath with ensemble temperature $T_{\text{E}}=1/\beta$. As shown in Fig. \ref{cigara}, the geometries with conical singularities correspond to geometries with different Hawking temperatures.
These are all possible fine-grained states, and so we should ensemble average over all possibilities that respect the setup.
Incorporating all possible geometries—whether or not they exhibit conical singularities within the Euclidean framework—is equivalent to integrating over all geometries represented in the free energy landscape illustrated in Fig.~\ref{Fig:FGibbs}.
In this way, the semi-classical physics and physics away from the saddles are both taken into account.

The density matrix can be expressed as follows
\be
\rho=\sum_{
\begin{tikzpicture}
\draw[thick] (-0.05,0) .. controls (-0.05,0) and (-0.175,0.0125) .. (-0.3,0.125);
\draw[thick] (-0.05,0.25) .. controls (-0.05,0.25) and (-0.175,1.9/8) .. (-0.3,0.125);
\draw[thick] (-0.05,0.125) ellipse (0.125/2 and 0.25/2);
\fill[red] (-0.3,0.125) circle (1pt);
\end{tikzpicture}
}
e^{-I_{\text{E}}[
\begin{matrix}
\begin{tikzpicture}
\draw[thick] (0.5/2,0) .. controls (0.5/2,0) and (0.25/2,0.025/2) .. (0,0.25/2);
\draw[thick] (0.5/2,0.5/2) .. controls (0.5/2,0.5/2) and (0.25/2,1.9/8) .. (0,0.25/2);
\draw[thick] (0.5/2,0.25/2) ellipse (0.125/2 and 0.25/2);
\fill[red] (0,0.25/2) circle (1pt);
\end{tikzpicture}
\end{matrix}
]}
\ket{
\begin{matrix}
\begin{tikzpicture}
\draw[thick] (0.5,0) .. controls (0.5,0) and (0.25,0.025) .. (0,0.25);
\draw[thick] (0.5,0.5) .. controls (0.5,0.5) and (0.25,1.9/4) .. (0,0.25);
\draw[thick] (0.5,0.25) ellipse (0.125 and 0.25);
\fill[red] (0,0.25) circle (1pt);
\end{tikzpicture}
\end{matrix}
}
\bra{
\begin{matrix}
\begin{tikzpicture}
\draw[thick] (-0.5,0) .. controls (-0.5,0) and (-0.25,-0.025) .. (0,-0.25);
\draw[thick] (-0.5,-0.5) .. controls (-0.5,-0.5) and (-0.25,-1.9/4) .. (0,-0.25);
\draw[thick] (-0.5,-0.25) ellipse (-0.125 and -0.25);
\fill[red] (0,-0.25) circle (1pt);
\end{tikzpicture}	
\end{matrix}
}.
\ee
The weight of different states, represented by the diagonal component of the density matrix, can be computed by evaluating the Euclidean action
\be
P[\begin{matrix}
\begin{tikzpicture}
\draw[thick] (0.5/2,0) .. controls (0.5/2,0) and (0.25/2,0.025/2) .. (0,0.25/2);
\draw[thick] (0.5/2,0.5/2) .. controls (0.5/2,0.5/2) and (0.25/2,1.9/8) .. (0,0.25/2);
\draw[thick] (0.5/2,0.25/2) ellipse (0.125/2 and 0.25/2);
\fill[red] (0,0.25/2) circle (1pt);
\end{tikzpicture}
\end{matrix}]
=
e^{-I_{\text{E}}[
\begin{matrix}
\begin{tikzpicture}
\draw[thick] (0.5/2,0) .. controls (0.5/2,0) and (0.25/2,0.025/2) .. (0,0.25/2);
\draw[thick] (0.5/2,0.5/2) .. controls (0.5/2,0.5/2) and (0.25/2,1.9/8) .. (0,0.25/2);
\draw[thick] (0.5/2,0.25/2) ellipse (0.125/2 and 0.25/2);
\fill[red] (0,0.25/2) circle (1pt);
\end{tikzpicture}
\end{matrix}
]}\,.\label{PP}
\ee
The expression above provides a formal representation of the density matrix; however, it is crucial to identify a parameter that characterizes the distinct states within the phase space. This parameter should serve as a label for the different Euclidean geometries illustrated in Fig.~\ref{cigara}.
Particularly, when performing ensemble averages, as depicted in Eq.~\eqref{aveF}, it is necessary to integrate over this parameter.
As already hinted in the free energy landscape, the parameter which we should use to label different states is the horizon radius $\rh$ corresponding to different horizon temperatures.
As can be seen from Fig.~\ref{cigara}, for Euclidean geometries the distance between the tip and boundary can be regarded as $|L-\rh|$.
The generalized free energy $F_\gen$ shown in Fig.~\ref{Fig:FGibbs} is a potential and one needs to consider perturbative kinetic energy on the potential to determine the canonical variable.
It was shown that the conjugate momentum should be $\dot{r}_{\text{h}}$ \cite{Li:2021vdp,Li:2022yti,Li:2022ylz}, and therefore it is natural to use $\rh$ as the canonical variable in defining the ensemble average.
Moreover, $\rh$ is typically regarded as the order parameter for describing black hole phase transitions, and using $\rh$ in the free energy landscape correctly captures the black hole phase transition.
With the parameter $\rh$ labeling different states in the phase space, the ensemble-averaged free energy shown in Eq.~\eqref{aveF} can be further written as
\be
\begin{split}
\ave{F}= \frac{1}{Z}~\tr(\rho F_{\gen}) = \frac{\int F_{\gen}~ e^{-I_{\text{E}} } d \rh }{\int e^{-I_{\text{E}} } d \rh}\,.\label{aveF0}
\end{split}
\ee
In terms of the normalized density matrix $\hat{\rho}=\rho/Z$, we have $\ave{F}=\tr(\hat{\rho} F_{\gen})$.
Then, following Eq.~\eqref{PP} the normalized probability distribution can be defined as
\be
\hat{P}[\begin{matrix}
\begin{tikzpicture}
\draw[thick] (0.5/2,0) .. controls (0.5/2,0) and (0.25/2,0.025/2) .. (0,0.25/2);
\draw[thick] (0.5/2,0.5/2) .. controls (0.5/2,0.5/2) and (0.25/2,1.9/8) .. (0,0.25/2);
\draw[thick] (0.5/2,0.25/2) ellipse (0.125/2 and 0.25/2);
\fill[red] (0,0.25/2) circle (1pt);
\end{tikzpicture}
\end{matrix}]
=\frac{P[\begin{matrix}
\begin{tikzpicture}
\draw[thick] (0.5/2,0) .. controls (0.5/2,0) and (0.25/2,0.025/2) .. (0,0.25/2);
\draw[thick] (0.5/2,0.5/2) .. controls (0.5/2,0.5/2) and (0.25/2,1.9/8) .. (0,0.25/2);
\draw[thick] (0.5/2,0.25/2) ellipse (0.125/2 and 0.25/2);
\fill[red] (0,0.25/2) circle (1pt);
\end{tikzpicture}
\end{matrix}]
}{Z}
=
\frac{e^{-I_{\text{E}}}}{\int e^{-I_{\text{E}}} d \rh}\,.
\ee
As we shall see later, the ensemble-averaged quantities contain all the physics of the black hole thermodynamics and some physics beyond the semi-classical limit.

\subsection{Black hole physics as a small $G_{\N}$ effect}
\label{sec:semi-classical}

The Euclidean action of the Kerr-AdS black hole is proportional to $1/G_{\N}\propto 1/l_{\text{pl}}^2$ as demonstrated in Sec.~\ref{sec22}, where $l_{\text{pl}}$ is the Planck length.
The semi-classical limit occurs when the Planck length $l_{\text{pl}}$ is much smaller than the AdS radius $L$.
Defining a dimensionless Newton's constant $\tilde{G}_{\N}$ following \cite{Cheng:2024hxh}, we have
\be
\tilde{G}_{\N}=\frac{l_{\text{pl}}^2}{L^2}\,.
\ee
The semi-classical regime corresponds to the condition $\tilde{G}_{\N}\ll 1$.
We would always assume that the size of the system $L$ is at a finite scale. Thus, it is harmless to drop the tilde sign in $\tilde{G}_{\N}$ and regard $G_{\N}\ll 1$ as the semi-classical limit.
In the context of the holographic duality, small $l_{\text{pl}}^{d-2}/L^{d-2}$ corresponds to large $N$ on the boundary. In such a situation, we have the classical geometry in the bulk.
Moreover, for a many-body system, $L\gg l_{\text{pl}}$ is the thermodynamic limit, where we should be able to recover the black hole thermodynamics. Away from the thermodynamic limit, there should not be any sharp phase transition.

\begin{figure}
    \centering
\subfloat[The Dirac delta distribution]{\includegraphics[width=0.49 \textwidth]{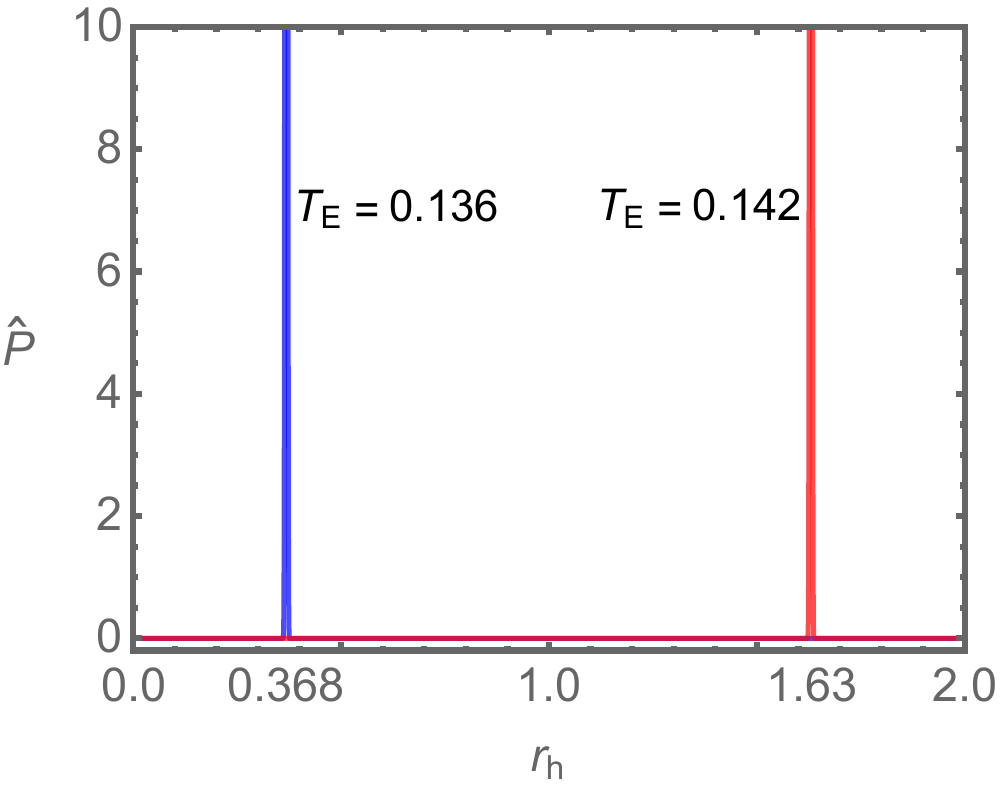}\label{delta}
}~~
\subfloat[The averaged free energy with $G_{\N}$=1/10000]{\includegraphics[width=0.51 \textwidth]{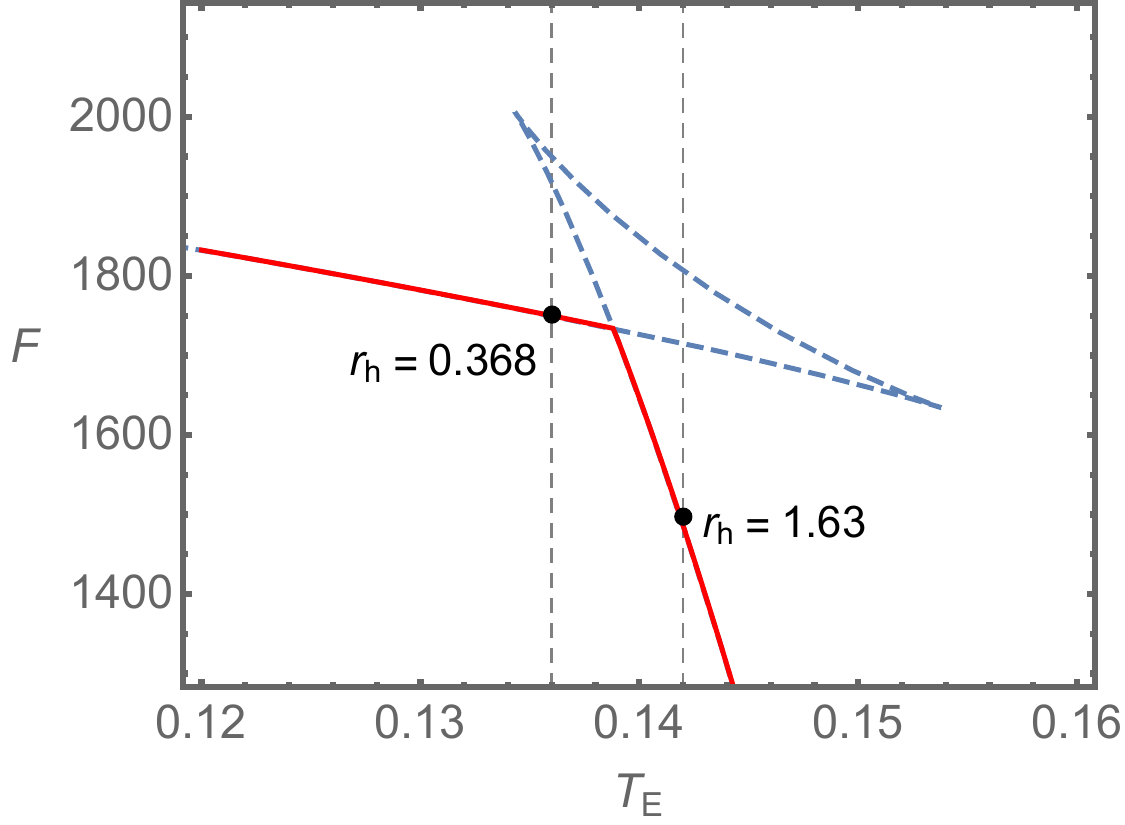}\label{G-10000}}
    \caption{(a) A demonstration of the probability distribution with small $G_{\N}$. For the Dirac delta distribution, the path integral (or ensemble average) only picks the saddle point contribution; (b) The semi-classical regime of the ensemble-averaged free energy $\ave{F}$ with $G_{\N}$=1/10000. We set angular momentum $J=0.05/G_{\N}$, and pressure $P=0.5~P_c$.}
    \label{smallGN}
\end{figure}

In the semi-classical limit, the probability distribution can be approximated by the Dirac delta function, as shown in Fig.~\ref{delta}.
When averaging over all the states in the phase space, the averaged free energy equals the saddle point value, allowing effects away from the saddles to be ignored.
Thus, we recover the black hole thermodynamics in the semi-classical limit.
As shown in Fig.~\ref{G-10000}, the averaged free energy defined in Eq.~\eqref{aveF0} is represented by the red curve in the semi-classical limit.
$G_{\N}$ is taken to be $1/10000$, and we can see that the ensemble-averaged free energy is extremely close to the classical Gibbs free energy depicted by the blue dashed line.
In the $G_{\N}\ll 1$ limit, we can observe that there is a sharp phase transition between large and small black holes.
This behavior encapsulates all aspects of black hole thermodynamics in the semi-classical limit.

Note that in the ensemble-averaged theory, the phase transition between small and large black holes is less artificial than the phase transition reviewed in Sec.~\ref{sec:landscape}. In traditional thermodynamics, we need to separately evaluate and compare the Gibbs free energy of different black hole states and manually select the one with the lowest free energy as the most stable. However, in our situation, the stable states are automatically selected by the probability distribution as demonstrated in Fig.~\ref{delta}.

\begin{figure}
    \centering
    \subfloat[Probability distribution with $G_{\N}$=1/100]{\includegraphics[width=0.48 \textwidth]{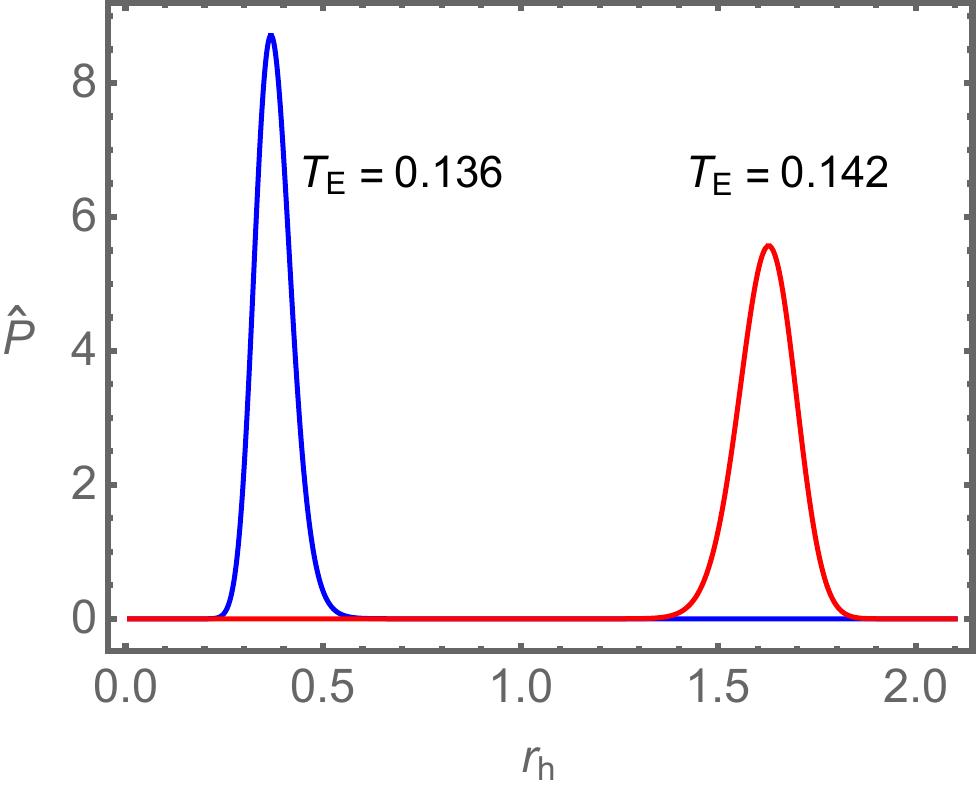}\label{largerGN-a}}~~~~
    \subfloat[The averaged free energy with $G_{\N}$=1/100]{\includegraphics[width=0.51 \textwidth]{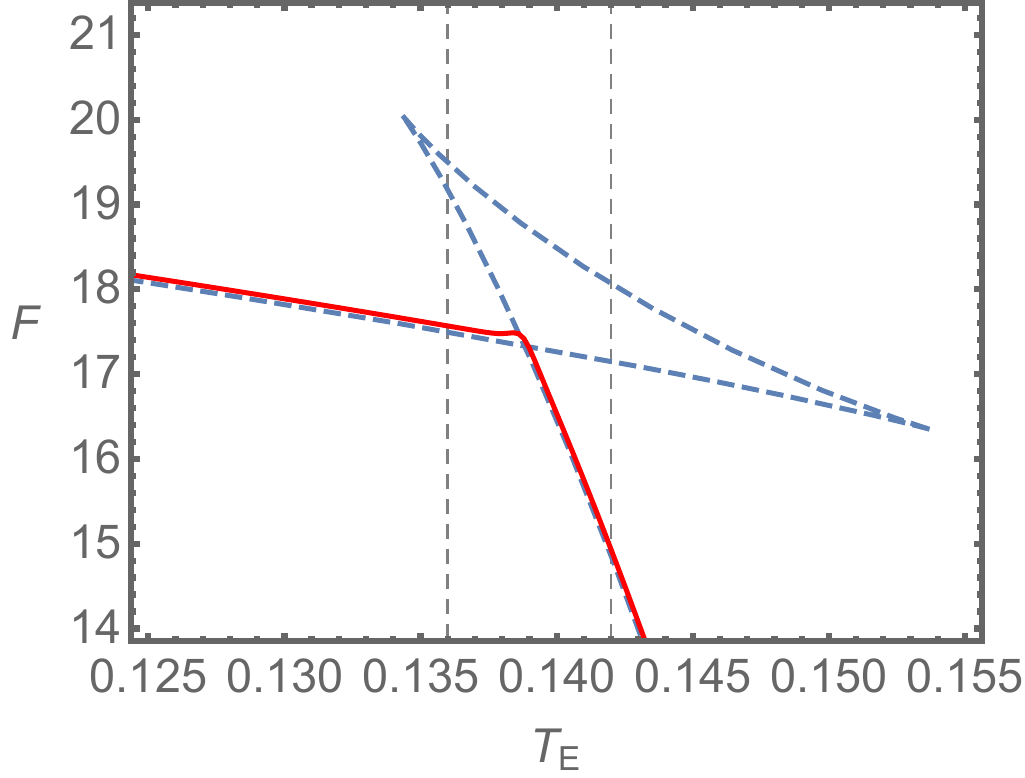}\label{largerGN-b}}
\\
    \subfloat[Probability distribution with $G_{\N}$=1/10]{\includegraphics[width=0.49 \textwidth]{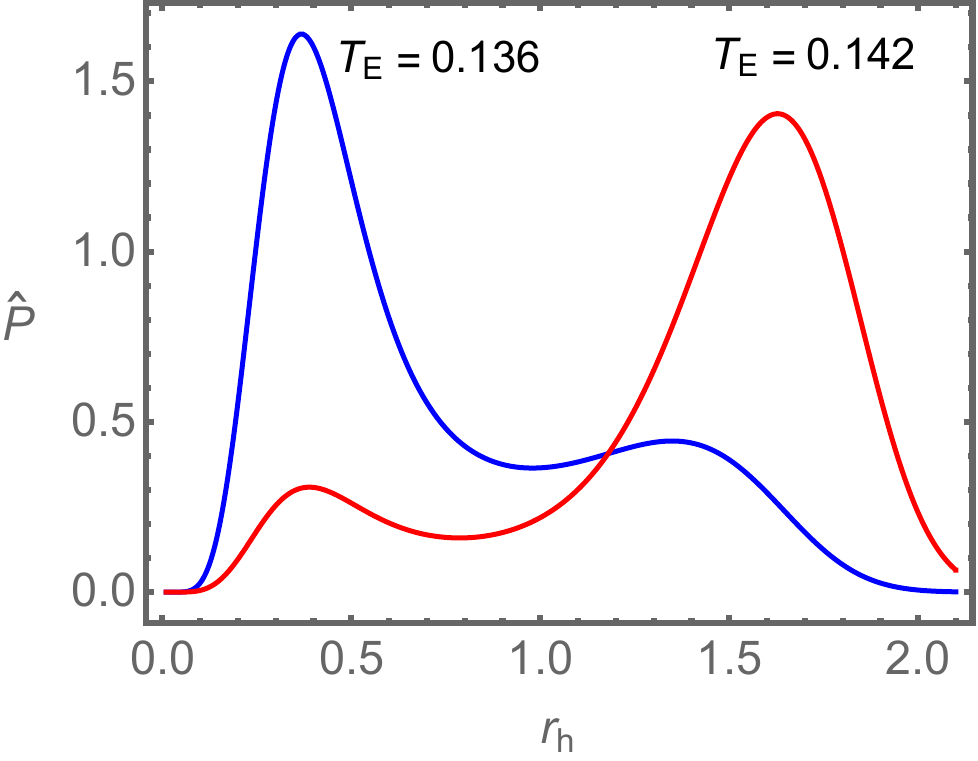}\label{largerGN-c}}~~~~
    \subfloat[The averaged free energy with $G_{\N}$=1/10]{\includegraphics[width=0.51 \textwidth]{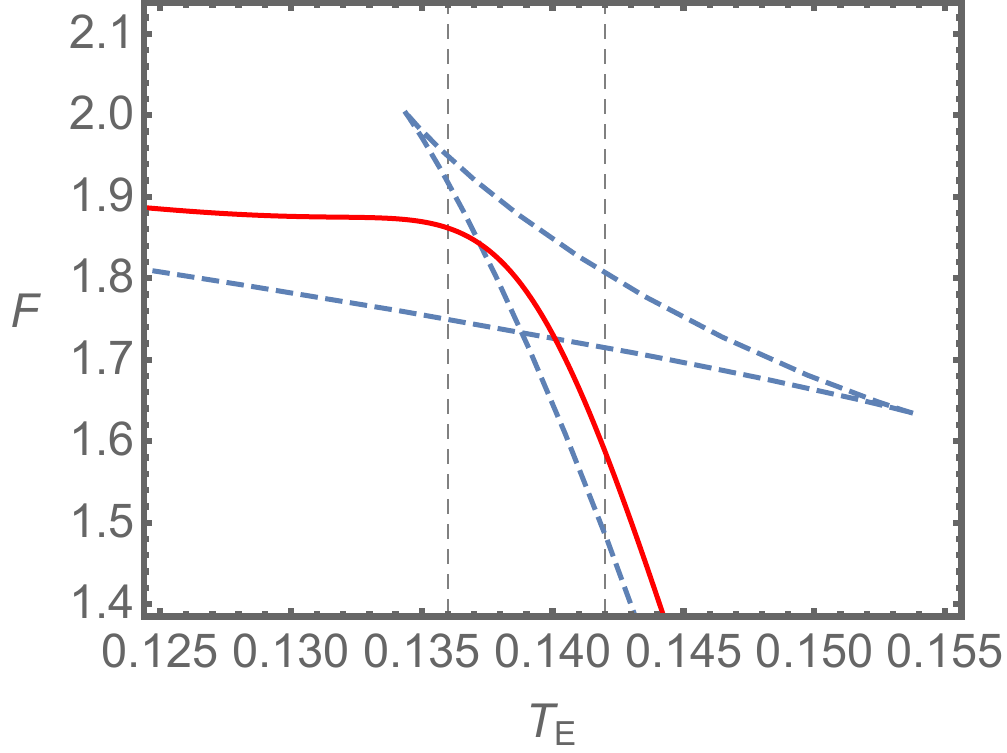}\label{largerGN-d}}
    \caption{Probability distribution $\hat{\rho}$ and the ensemble-averaged free energy $\ave{F}$ with different values of $G_{\N}$. We have chosen $J=0.05/G_{\N}$, and $P=0.5~P_c$. For different ensemble temperatures $T_\text{E}$, we have different configurations of $\hat{\rho}$. For $T_\text{E}=0.136$, the system can be regarded as in the small black hole phase. For $T_\text{E}=0.142$, the system is in the large black hole phase. The ensemble-averaged free energies have deviations away from the black hole Gibbs free energy.}
    \label{largerGN}
\end{figure}

For larger $G_{\N}$, the normalized probability distribution $\hat{P}$ are plotted in Figs. \ref{largerGN-a} and \ref{largerGN-c}, with different values of $G_{\N}$. As can be seen from the figures, for $G_{\N}=1/100$, the classical black hole and neighboring states in the free energy landscape are the most important ones. When averaging over all the states in the landscape, we get the corrected free energy as can be seen from Fig.~\ref{largerGN-b}. The probability distribution can be approximated by the Gaussian distribution, for $G_{\N}=1/100$, and we will see the leading-order corrections analytically later.
For $G_{\N}=1/10$, the contributions from different saddles can not be ignored, and the averaged free energy would significantly deviate from the black hole Gibbs free energy, as shown in Fig. \ref{largerGN-d}.
Although the state with radius $\rh$ around $0.368$ is the most important state for ensemble temperature $T_\text{E}=0.136$ as can be seen from \ref{largerGN-c}, there is another peak at $\rh=1.63$. Similarly for $T_\text{E}=0.142$, in addition to the peak at $\rh=1.63$ there is also a smaller peak in the weight distribution at $\rh=0.368$.
From Figs. \ref{largerGN-a} and \ref{largerGN-c}, we can still see some sign of phase transition. For $T_\text{E}=0.136$, the system can be more or less regarded as in the small black hole phase. Or one can say that the black hole state with a small horizon radius has a larger weight. While $T_\text{E}=0.136$, the system is in a small black hole phase.
However, there is no sharp transition of phase because we are away from the thermodynamical limit (semiclassical limit).

The Gibbs free energy is proportional to $1/G_{\N}$, and the shape of the classical Gibbs free energy (the swallowtail diagram) is not affected by $G_{\N}$. However, the shape of the probability distribution, as shown in Figs. \ref{delta}, \ref{largerGN-a} and \ref{largerGN-c}, can be significantly influenced by the value of $G_{\N}$.
For small $G_{\N}$, the distribution can be approximated by the Dirac delta function. For relatively larger $G_{\N}$, it can be approximated by a Gaussian distribution as can be seen from Fig. \ref{largerGN-a}. For even larger $G_{\N}$, additional non-classical contributions must be considered.
The deviations from the classical saddles should be regarded as quantum effects, and we examine these corrections in detail later.

\section{Subleading-order corrections to black hole thermodynamics}\label{sec:subleading}

Based on the insights from the preceding section, we know that black hole thermodynamics emerges within the semi-classical limit from the ensemble average.
More specifically, as shown in Fig. \ref{G-10000} we can derive the correct black hole Gibbs free energy $F_\text{Gibbs}$ from the ensemble-averaged free energy $\ave{F}$ in the small $G_{\N}$ limit. The Gibbs free energy
$F_\text{Gibbs}$ is proportional to $1/G_{\N}$ as reviewed in Eq. \eqref{FGibbs}. Consequently,
as a function of $G_{\N}$, the ensemble-averaged free energy $\ave{F}$ must incorporate $F_\text{Gibbs}$ as the leading-order contribution.
Then the primary task of this section is twofold: to verify whether the leading-order contribution is $F_\text{Gibbs}$, and to determine the subleading and even higher-order corrections to black hole thermodynamics as we expand $\ave{F}$ in terms of $G_{\N}$.

With fixed angular momentum $J$ and pressure $P$, the ensemble-averaged free energy $\ave{F}$ is a function of $G_{\N}$ and ensemble temperature $T_\text{E}$.
We take $J=0.05/G_{\N}$ and $P=0.5 P_c$ in this section. There are two methods available for determining the $G_{\N}$ dependence of $\ave{F}$.
First, we can employ numerical techniques to fit the coefficients of the $G_{\N}$ expansion. Alternatively, we can utilize a Gaussian distribution to approximate the probability distribution for small $G_{\N}$, as illustrated in Fig. \ref{largerGN-a}. This approach allows us to derive an analytical expression for both the leading-order contribution and the subleading correction.
We use these two different methods and compare the results in the next two subsections.

%
%

\subsection{Numerical methods}
\begin{figure}
    \centering
    \subfloat[The leading-order coefficient]{\includegraphics[width=0.55 \textwidth]{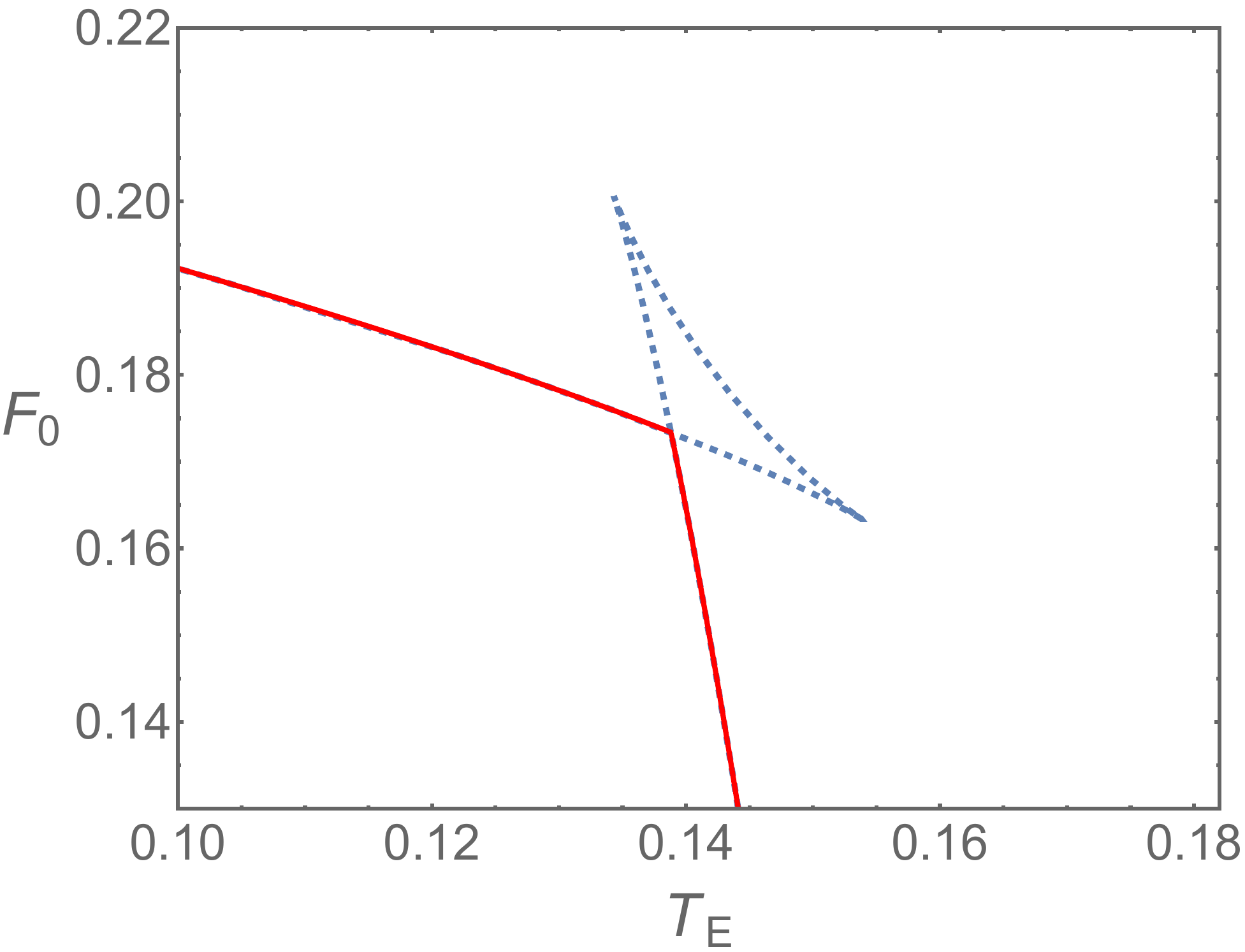}\label{fig.GNexpand-leading}}
\\
    \subfloat[The subleading-order coefficient]{\includegraphics[width=0.45 \textwidth]{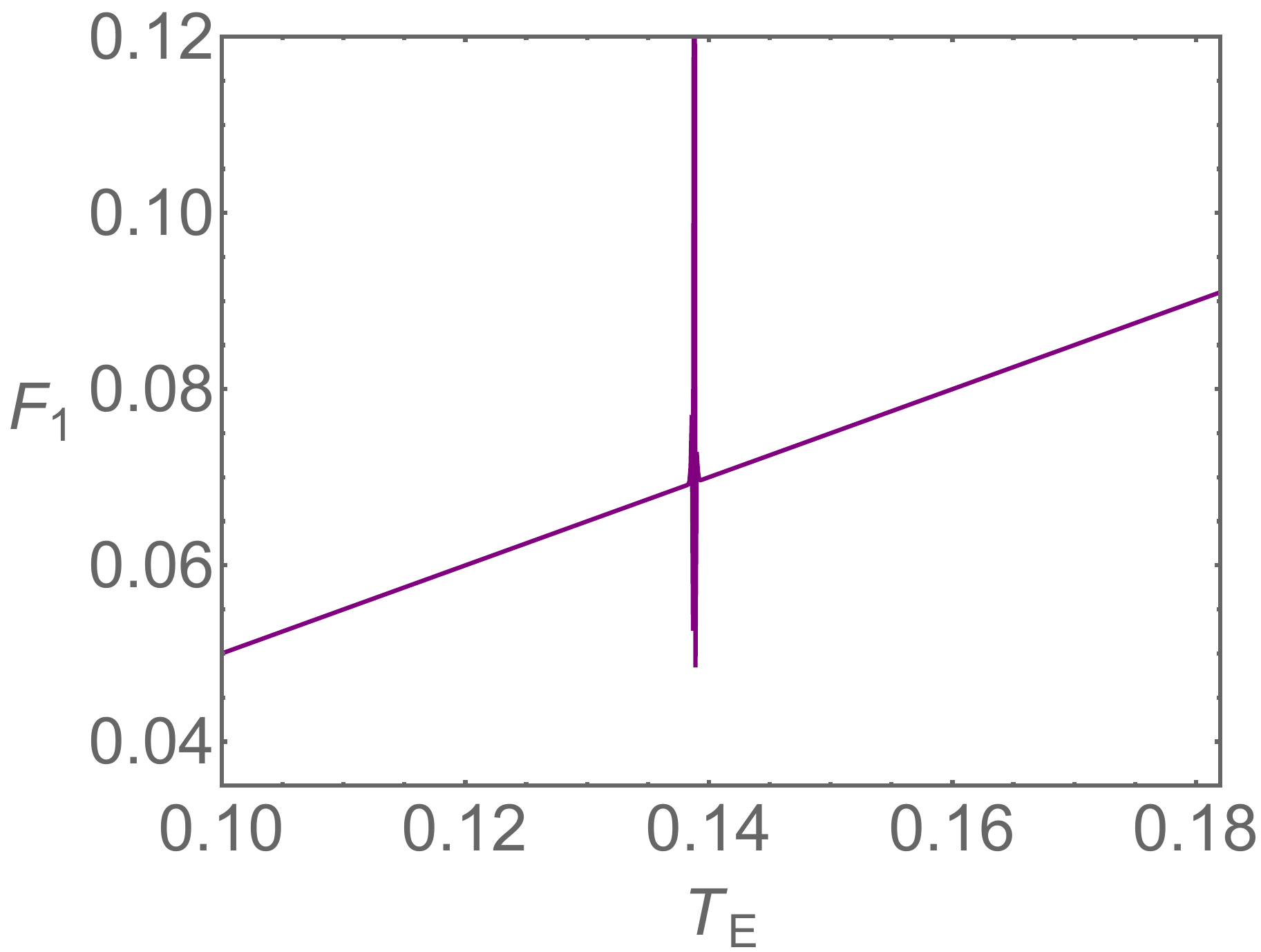}\label{GNexpand-b}}~~~~
    \subfloat[The subsubleading-order coefficient]{\includegraphics[width=0.45 \textwidth]{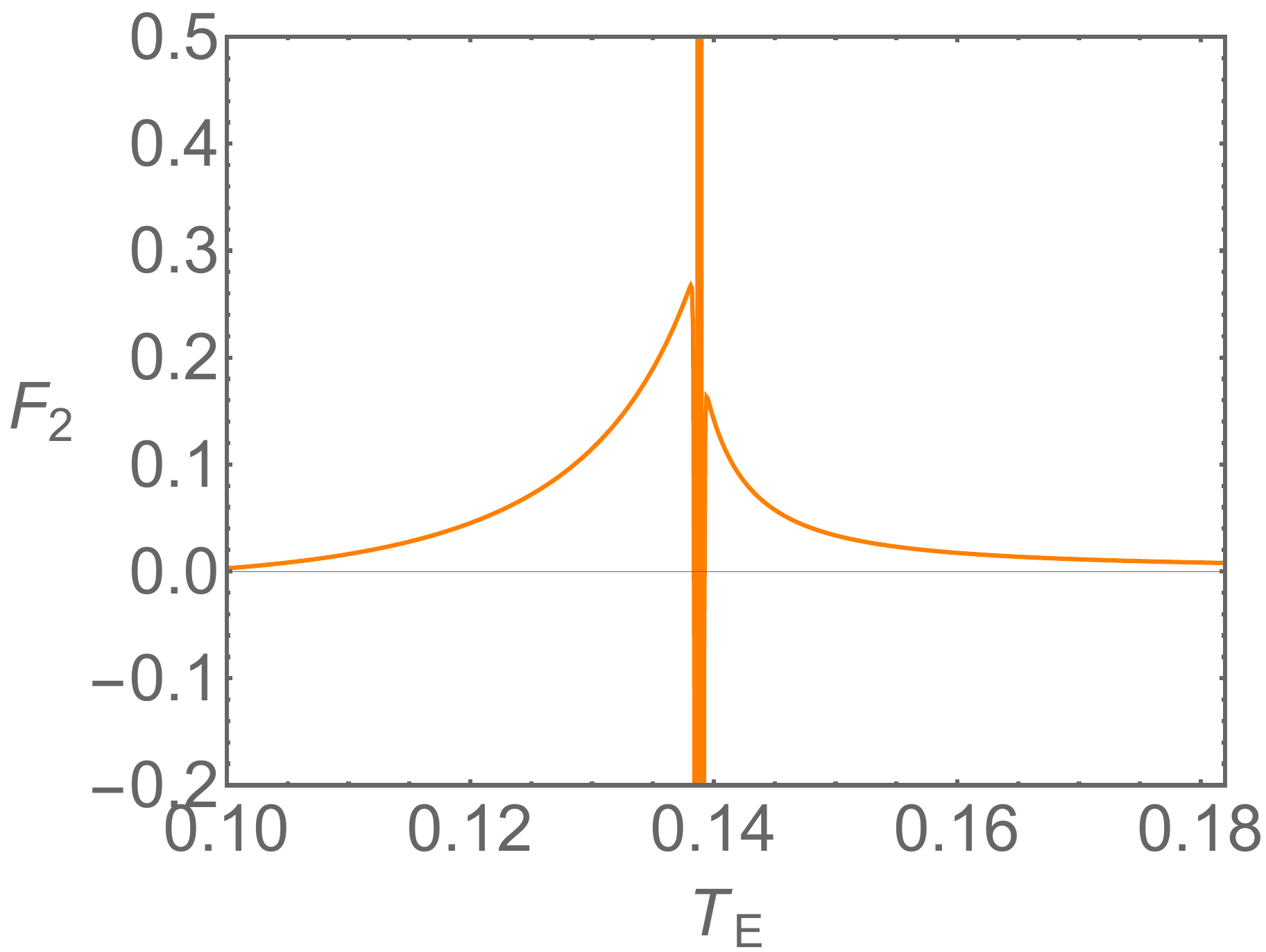}\label{GNexpand-c}}
    \caption{(a) The red line represents the leading order contribution $F_0$. It aligns perfectly with the black hole Gibbs free energy $F_{\text{Gibbs}} \cdot G_{\N}$ shown by the dashed line.
    (b) The purple line represents the subleading contribution of the averaged free energy $\ave{F}$. The subleading contribution
    $F_1$ has a linear dependence on $T_\text{E}$, which can be fitted as $F_2=0.500454 ~T_{\text{E}}+5.09608\times 10^{-6}$.
    (c) The orange line $F_2$ represents the sub-subleading contribution of the averaged free energy $\ave{F}$.}
    \label{GNexpand}
\end{figure}

By utilizing the expression for the ensemble-averaged free energy~\eqref{aveF0}, we can compute $\ave{F}$ for any specified value of $G_{\N}$. Through numerical methods, we can generate a dataset of $\ave{F}$ corresponding to various values of $G_{\N}$ and $T_\text{E}$. This process allows us to plot the red curve representing $\ave{F}$ in Figs.~\ref{G-10000},~\ref{largerGN-b}, and \ref{largerGN-d}. Moreover, we can expand our analysis to include additional values of $G_{\N}$ beyond those presented in the figures, enabling a more comprehensive fit of the $G_{\N}$ dependence of $\ave{F}$.

As previously discussed, for small $G_{\N}$, the leading-order contribution is proportional to $1/G_{\N}$. Consequently, we can conduct a polynomial expansion of $G_{\N}$, with the leading-order term scaling as $1/G_{\N}$.
The expression can be expressed as
\begin{equation}
    \ave{F}=\frac{F_0}{G_{\N}} +F_1+F_2 G_{\N}+F_3 G_{\N}^2+\cdots.
\end{equation}
Then the main task is to determine the appropriate coefficients of the above expression: $F_0$, $F_1$, $F_2$, and so forth. For small values of $G_{\N}$, specifically  $G_{\N}<1/500$ in our case, only a limited number of leading terms are necessary to achieve a good fit for the ensemble-averaged free energy $\ave{F}$.

The results for $F_0$, $F_1$, and $F_2$ are illustrated in Fig. \ref{GNexpand}.
The term $F_0/G_{\N}$ is the leading-order contribution, which is shown to match exactly with the black hole Gibbs free energy.
Figs. \ref{GNexpand-b} and \ref{GNexpand-c} illustrate the subleading and sub-subleading contributions, respectively. For the subleading contribution, $F_1$ shows a linear dependence with the ensemble temperature $T_\text{E}$, except at the phase transition point. This result exactly matches the analytical result presented later. With regard to the sub-subleading contribution presented in Fig. \ref{GNexpand-c}, $F_2$ can also be fitted using data away from the transition point, following the methodology applied to the subleading-order term.
For small $G_{\N}$, for example $G_{\N}<1/500$, one can easily show that summing over the leading-order, subleading-order, and subsubleading-order contributions is a good approximation of $\ave{F}$, because the higher-order corrections are suppressed by $G_{\N}$ and can be ignored.

In summary, we can conclude that, within the context of the small
$G_{\N}$ expansion, the averaged free energy $\ave{F}$  can be expressed as follows:
\be
\ave{F}=F_\text{Gibbs}+F_1\cdot G_{\N}^0+F_2\cdot G_{\N}^1+\mathcal{O}(G_{\N}^2)\,,
\ee
where $F_\text{Gibbs}$ is the leading-order contribution that proportional to $1/G_{\N}$,  $F_1$ and $F_2$ are explicitly shown in Figs. \ref{GNexpand-b} and \ref{GNexpand-c}. We observe that black hole thermodynamics emerges in the small $G_{\N}$ limit, while for finite $G_{\N}$, the free energy experiences corrections, as illustrated in Fig.~\ref{GNexpand}.


\subsection{Analytical methods}

\begin{figure}[H]
    \centering
    \subfloat[$T_\text{E}$=0.136]{\includegraphics[width=0.48 \textwidth]{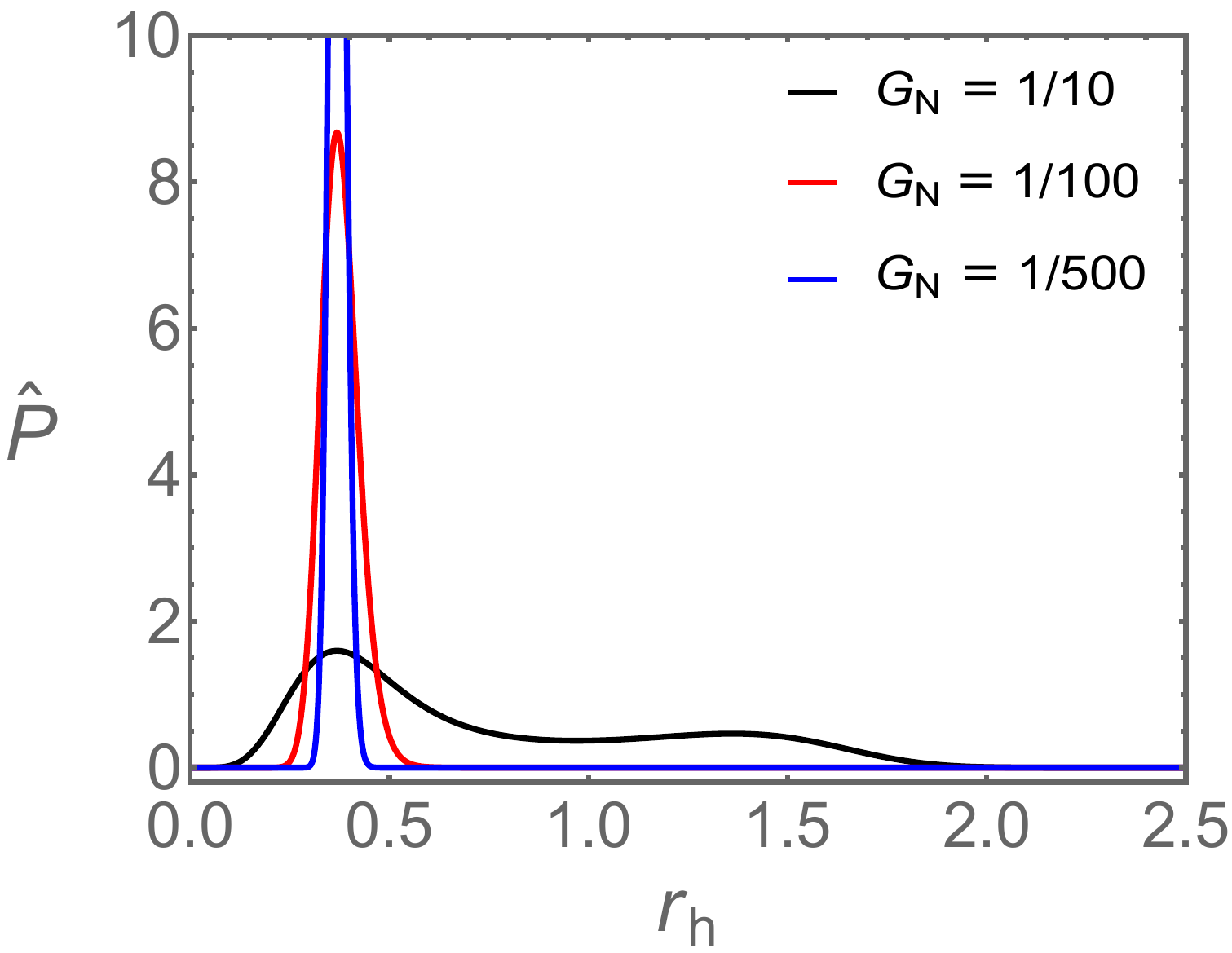}}~~~~
    \subfloat[$T_\text{E}$=0.142]{\includegraphics[width=0.48 \textwidth]{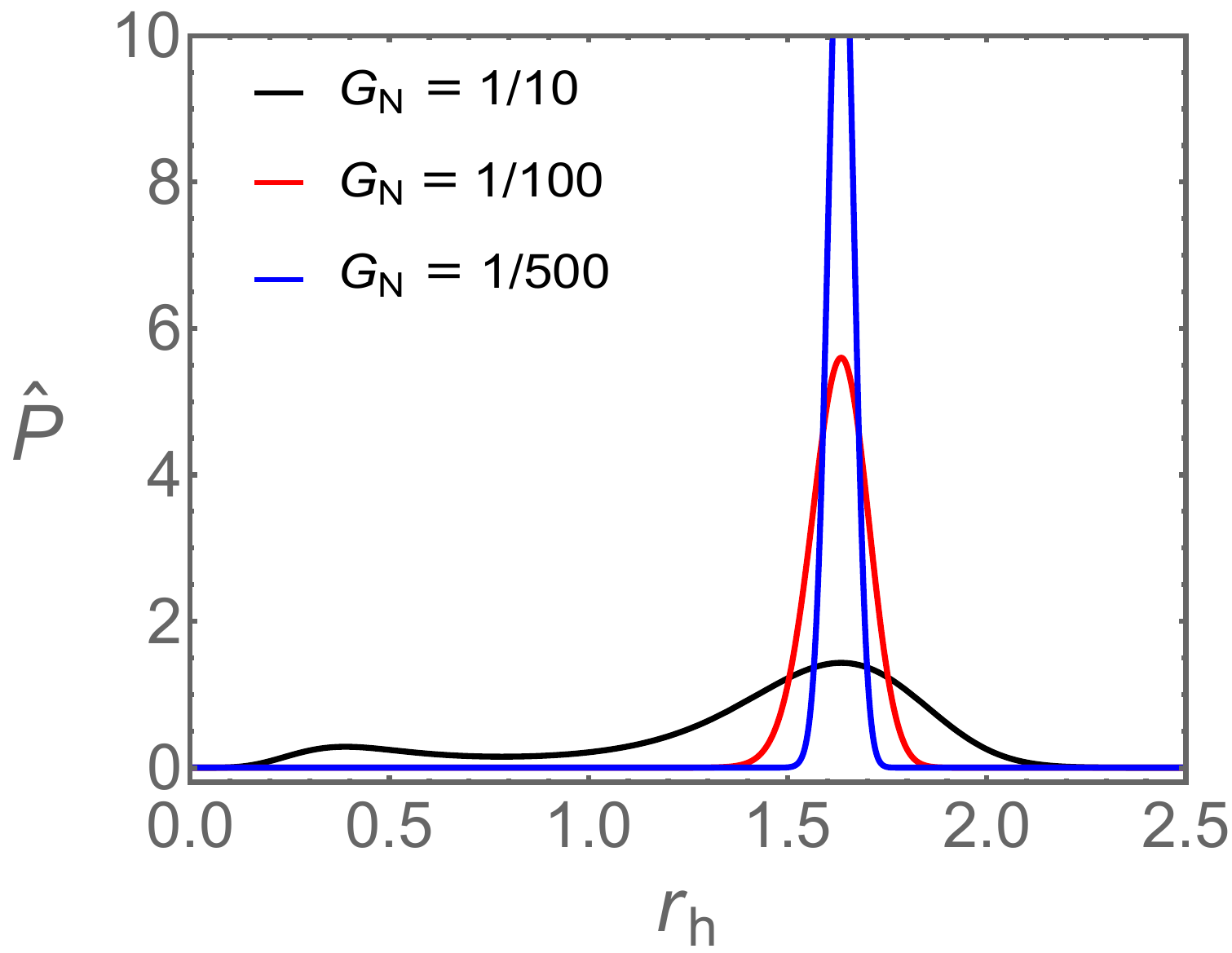}}
    \caption{The normalized probability distribution functions with different $G_{\N}$. The ensemble temperatures are set to $T_\text{E}=0.136$ and $T_\text{E}=0.142$ in (a) and (b), respectively. }
    \label{fig:Prob}
\end{figure}

We can also conduct an analytical examination of the leading- and subleading-order contributions to the ensemble-averaged free energy.
As demonstrated in Fig.~\ref{fig:Prob}, the shape of the probability distribution varies with different values of $G_{\N}$.
When the ensemble temperature is fixed, the peaks of the curves align at the same black hole state $\rh$.
When $G_{\N}=1/10$, the probability of other geometries, especially the large black hole phase, can not be ignored as shown by the black curves in Fig.~\ref{fig:Prob}.
For small $G_{\N}$ (blue and red curves), the other peak is negligible and we have a rather sharp and narrow probability distribution.
In this case, the probability distribution can be approximated by a Gaussian distribution with expectation value at $\rh=r_\h$, where $r_\h$ is defined as the horizon radius when the Hawking temperature $T_{\text{H}}$ equals to the ensemble temperature $T_\text{E}$. As for larger $G_N$, even though the whole probability
distribution does not resemble a Gaussian, the vicinity of the maximum of $\hat{P}$ can be approximated by the Gaussian distribution.
This means that we can expand the Euclidean action up to the second order and directly evaluate the averaged free energy.
Expanding around $\rh=r_\h$, the Euclidean action \eqref{IE} can be written as
\bea
I^{\delta}_\text{E} &=&\frac{\beta_\h r_\h}{2 G_{\N} \Xi^2}\left(1+\frac{a^2}{r_\h^2}\right)\left(1+\frac{r_\h^2}{L^2}\right)-\frac{\pi\left(r_\h^2+a^2\right)}{G_{\N} \Xi}\nonumber\\
&~&+ \left[\frac{\beta_\h}{2G_{\N} \Xi^2 r_\h}\left( \frac{a^2}{r_\h^2}+\frac{3r_\h^2}{L^2}\right)-\frac{\pi}{G_{\N} \Xi} \right]\delta^2\,,\label{Idelta}
\eea
where $\delta$ is defined as $\delta=\rh-r_\h$. The first-order term is zero because we have $\partial_{\rh}I_{\text{E}}=0$ near extreme points.
The generalized free energy can be expanded as
\bea
F^{\delta}_{\text{gen}} &=& \frac{r_\h}{2 G_{\N} \Xi^2}\left(1+\frac{a^2}{r_\h^2}\right)\left(1+\frac{r_\h^2}{L^2}\right)-\frac{\pi T_\h\left(r_\h^2+a^2\right)}{G_{\N} \Xi}\nonumber\\
&~&+ \left[\frac{1}{2G_{\N} \Xi^2 r_\h}\left( \frac{a^2}{r_\h^2}+\frac{3r_\h^2}{L^2}\right)-\frac{\pi T_\h}{G_{\N} \Xi} \right]\delta^2\,,
\eea
Note that the ensemble temperature $T_\text{E}$ equals the Hawking temperature $T_\h$ when $\rh=r_\h$.
Now the averaged free energy can be approximated by
\be
\ave{F}= \frac{1}{Z}~\tr(\rho F_{\gen}) \approx
\frac{\int_{-5\sigma}^{5\sigma} F^{\delta}_{\gen}~ e^{-I^{\delta}_{\text{E}} } d \delta }{\int_{-5\sigma}^{5\sigma} e^{-I^{\delta}_{\text{E}} } d \delta}\,.\label{GaussF}
\ee
The integral is integrated from $-5\sigma$ to $5\sigma$, where $\sigma$ is the standard deviation of the Gaussian distribution. Denoting the coefficient in front of $\delta^2$ in Eq.~\eqref{Idelta} as $I^{(2)}_{\text{E}}$, the standard deviation can be expressed as
\be
\sigma =\frac{1}{\sqrt{2 I^{(2)}_{\text{E}}}}\,.
\ee

The Gaussian integral \eqref{GaussF} can be directly evaluated, and we get
\bea
\ave{F} &\approx& \frac{1}{4G_{\N} \Xi r_\h}\left[a^2\Big(1-\frac{r_\h^2}{L^2}\Big)+\frac{2(a^2+r_\h^2)(L^2-r_\h^2)}{L^2-a^2}-r_\h^2\Big(1+3\frac{r_\h^2}{L^2}\Big)\right]\nonumber\\
&~& +\frac{T_\h}{2}-\frac{5 e^{-25/2}}{ \sqrt{2\pi}\text{erf}(\frac{5}{\sqrt{2}})}T_\h\,.\label{anaF}
\eea
The first term in Eq.~\eqref{anaF} proportional to $1/G_{\N}$ is exactly the Gibbs free energy of the Kerr-AdS black hole.
The second term $T_\h/2$ is the subleading-order contribution of the $G_{\N}$ expansion, which exactly matches the result from the numerical methods. This can be directly seen from Fig. \ref{GNexpand-b}. The agreement between the numerical and analytical results demonstrates that the Gaussian distribution is a valid approximation for the distribution function at small $G_{\N}$.
The error term in Eq.~\eqref{anaF}, arising from the integration over the interval $(-5\sigma,5\sigma)$, is $7\times10^{-6}\cdot T_\h$, and can be considered negligible in comparison to $T_\h/2$.

For the Kerr-AdS black hole, there are two phases—the small and large black hole phases—as shown in Fig. \ref{fig:Prob}.
Here we only care about the situation below the critical pressure.
The location of the peak shifts as the ensemble temperature changes from $T_\text{E}=0.136$ to $T_\text{E}=0.142$, indicating a change in the classical black hole radius. The black hole horizon radius does not change continuously with the ensemble temperature; instead, it changes abruptly at a transition temperature, signifying a phase transition from a small black hole to a large black hole.
In different phases, we should expand around different values of $r_\h$, either $r_\h=0.368$ or $r_\h=1.63$ for the chosen parameters.
Although the black hole free energy can vary in different phases, the subleading contribution is always $T_\h/2$.
This conclusion aligns with the result in Fig. \ref{GNexpand-b}: whether the temperature is lower or higher than the critical temperature, the subleading-order contribution $F_2$ is always a linear function of the temperature with a slope $1/2$.

\begin{figure}
    \centering
    \includegraphics[width=0.55 \textwidth]{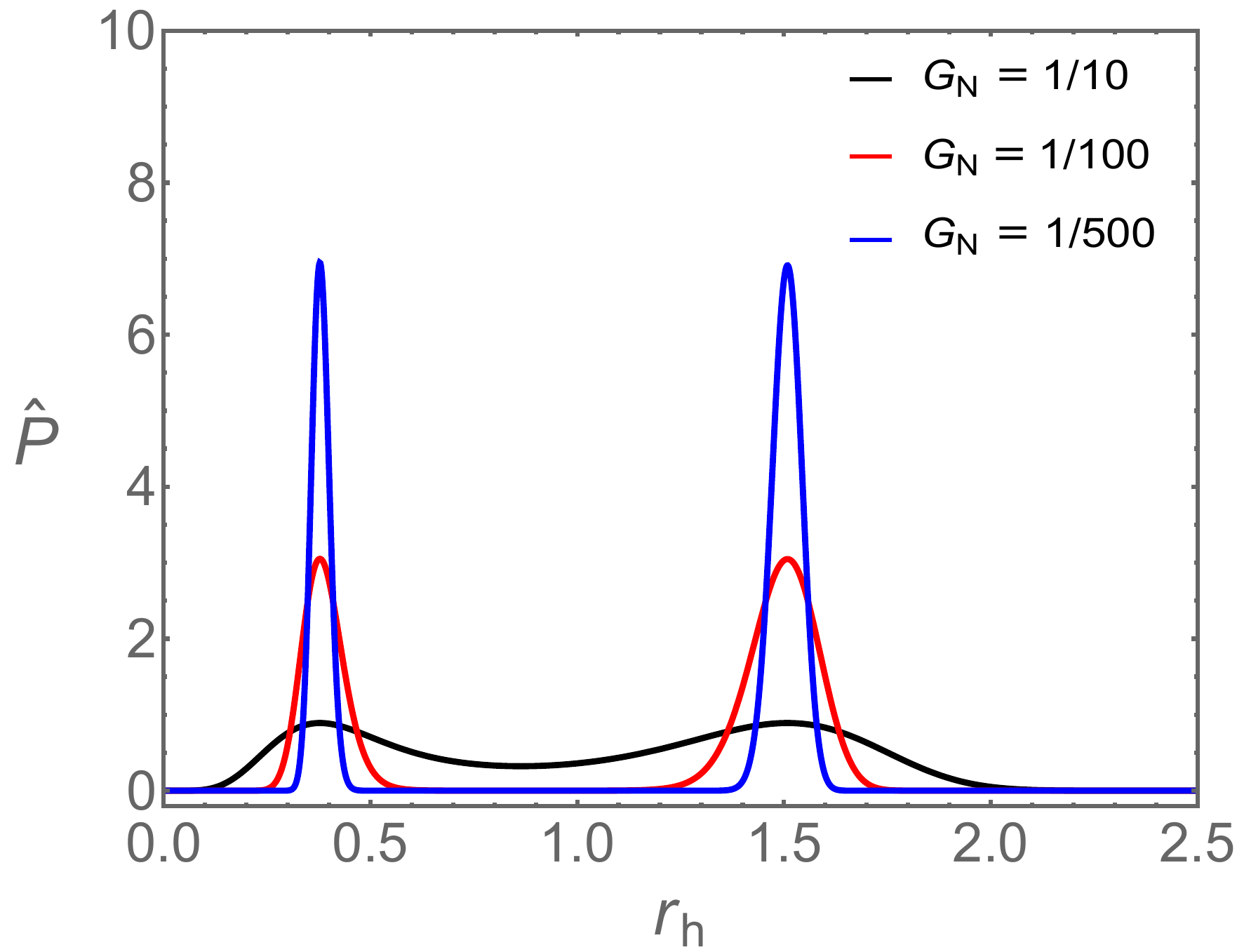}
    \caption{At the transition temperature, the heights of the two peaks are the same. The contributions of the two saddles to the path integral are comparable.}
    \label{Prob_c}
\end{figure}

As a brief comment on the subleading-order contribution, we notice that there is a violent oscillation behavior near the transition point as shown in Fig. \ref{GNexpand-b}.
This can be explained by Fig. \ref{Prob_c}, where the probability distribution is illustrated for the situation that the ensemble temperature equals the transition temperature.
At the transition temperature, the two peaks have nearly the same contribution, making it unsuitable to approximate the original probability distribution with a Gaussian distribution.
That is the reason why we have violent oscillation behaviors near the transition points.
Nevertheless, since the distribution function exhibits rapid changes due to the exponential term, the Gaussian approximation should be applicable once we move away from the phase transition point.

\subsection{A universal subleading-order contribution for all asymptotic AdS black holes}

For Schwarzschild-AdS and RN-AdS black holes~\cite{Cheng:2024hxh}, the ensemble-averaged free energy can be defined similarly.
In this context, the classical Gibbs free energy emerges as the leading-order contribution, while half of the Hawking temperature serves as the subleading-order contribution.
In the previous subsection, we established that for the Kerr-AdS black hole, despite the increased complexity of the scenario, the ensemble-averaged free energy exhibits behavior akin to that observed in the Schwarzschild-AdS and RN-AdS black holes~\cite{Cheng:2024hxh}.
All cases being studied suggest that this behavior should be universal for all asymptotic AdS black holes. Consequently,
we can assert the following universal claim:\\

\noindent \textbf{Claim}: \textit{For asymptotic AdS black holes, the leading-order contribution of the ensemble-averaged free energy is the classical Gibbs free energy, and the subleading correction to the Gibbs free energy due to finite $G_{\N}$ effect is always half of the Hawking temperature.}\\

\noindent \textbf{Proof}:\\
For any asymptotic AdS black hole with conical singularity, the Euclidean action can be evaluated and denoted as $I_\text{E}(\rh)$. The action can depend on other parameters such as conserved charges $Q_i$, pressure $P$, angular momentum $J$, and so forth, which are not written explicitly in $I_\text{E}(\rh)$.
Following the arguments from Sec.~\ref{sec: average}, the normalized probability distribution can be denoted as
\be
\hat{P}=\frac{e^{-I_E}}{Z}\,.
\ee
As the on-shell configuration, the classical solution with horizon radius $r_\h$ should always correspond to the local extremum of both the probability distribution and the generalized free energy defined as $F_\text{gen}=I_{\text{E}}/\beta$.
So one can expand $I_{\text{E}}$ around the local extremum up to the second order of $\delta=\rh-r_\h$.
For the Kerr-AdS case, this means that we use a Gaussian distribution to approximate the vicinity of the maximum of $\hat{\rho}$.
For small $G_{\N}$ we can use the Gaussian integral to approximate the whole distribution; while for larger $G_{\N}$ we can integrate over the neighborhood of the extreme point from $-5\sigma$ to $5\sigma$.

Expanding the Euclidean action near the neighborhood of $\rh=r_\h$, we have
\be
I_{\text{E}}(\rh)\approx I_{\text{E}}(r_\h)+I'_{\text{E}}(r_\h)\delta+\frac{I''_{\text{E}}(r_\h)}{2}\delta^2=I_{\text{E}}(r_\h)+\frac{I''_{\text{E}}(r_\h)}{2}\delta^2\,,
\ee
where the prime means taking derivative with respect to $\rh$. $I'_{\text{E}}$ is always zero because we are expanding everything near the extreme points. The corresponding generalized free energy can be written as
\be
F_{\text{gen}}=\frac{I_{\text{E}}}{\beta}\approx \frac{I_{\text{E}}(r_\h)}{\beta}+\frac{I''_{\text{E}}(r_\h)}{2\beta}\delta^2\,.
\ee
Thus, the ensemble-averaged free energy $\ave{F}$ can be expressed as follows:
\begin{equation}
    \begin{split}
      \ave{F}&=\tr(\hat{\rho}F_{\text{gen}})\\
      &=\frac{1}{Z}\int_{-5\sigma}^{5\sigma}\left(\frac{I_{\text{E}}(r_\h)}{\beta}+\frac{I''_{\text{E}}(r_\h)}{2\beta}\delta^2\right)
      \exp\left[ -I_{\text{E}}(r_\h)-\frac{I''_{\text{E}}(r_\h)}{2}\delta^2\right]d\delta\,,
    \end{split}
\end{equation}
with the partition function
\be
Z=\int_{-5\sigma}^{5\sigma}\exp\left[{-I_{\text{E}}(r_\h)-\frac{I''_{\text{E}}(r_\h)}{2}\delta^2}\right]d\delta\,,
\ee
and the standard deviation
\be
\sigma=\frac{1}{\sqrt{I''_{\text{E}}(r_\h)}}\,.
\ee
The Gaussian integral can be evaluated directly, yielding the ensemble-averaged free energy as follows:
\be
\ave{F}=\frac{I_{\text{E}}(r_\h)}{\beta}+\frac{T_\h}{2}-\frac{5 e^{-25/2}}{ \sqrt{2\pi}\text{erf}(\frac{5}{\sqrt{2}})}T_\h\,.
\ee
The term related to the error function arises due to the domain of integration $(-5\sigma,5\sigma)$, and is much smaller than ${T_\h}/{2}$.
Hence, the error function term can be ignored and we are left with the classical black hole free energy and a linear term in $T_\h$.

Thus, we have established a universal claim for asymptotic AdS black holes: the ensemble-averaged free energy comprises the classical Gibbs free energy as the leading-order contribution, along with a term that is half of the Hawking temperature as the subleading correction. It is intriguing to explore more black hole cases to verify the claim or find counter-examples in different situations.

\subsection{Quantum-corrected black hole thermodynamics} \label{sec:loop}

In the previous part, an important conclusion is that the subleading-order correction to the black hole Gibbs free energy is proportional to half of the Hawking temperature.
The existence of this correction arises from the inclusion of off-shell geometries with conical singularities in the phase space.
In terms of the probability distribution of the states, we incorporate quantum effects beyond the classical black hole physics by replacing a single saddle point with a Gaussian distribution centered around of the saddle point.
While the Gaussian distribution serves as an approximation for the distributions with small $G_{\N}$ depicted in Fig.~\ref{fig:Prob}, it is good enough to capture the subleading-order contribution.
By incorporating effects beyond the classical saddle point, we suggest a new quantum-corrected thermodynamics for black holes.

\begin{figure}
    \centering
     \subfloat[$G_{\N}$=1/50]{
    \includegraphics[width=0.48\linewidth]{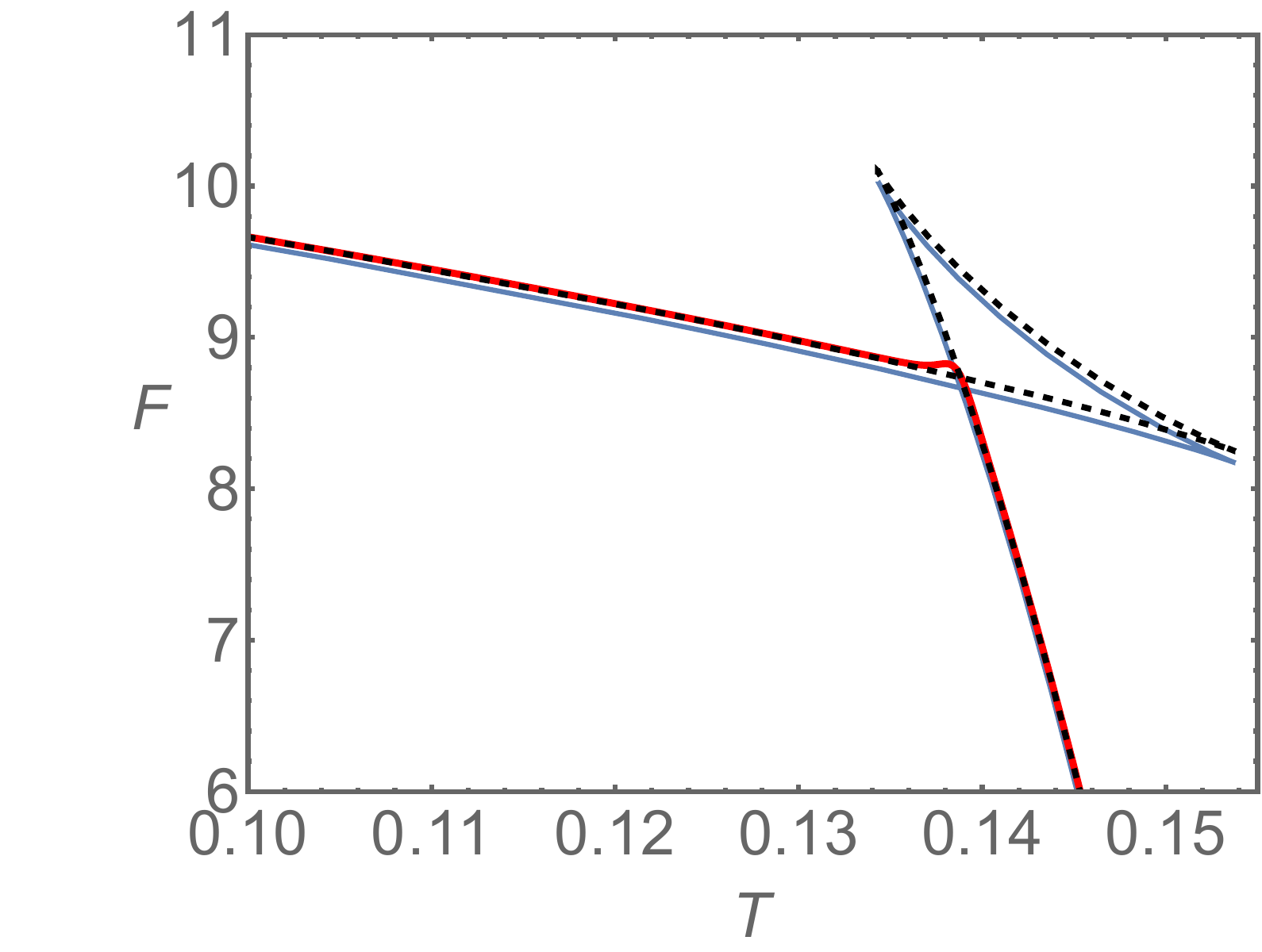}}
     \subfloat[$G_{\N}$=1/10]{
    \includegraphics[width=0.48\linewidth]{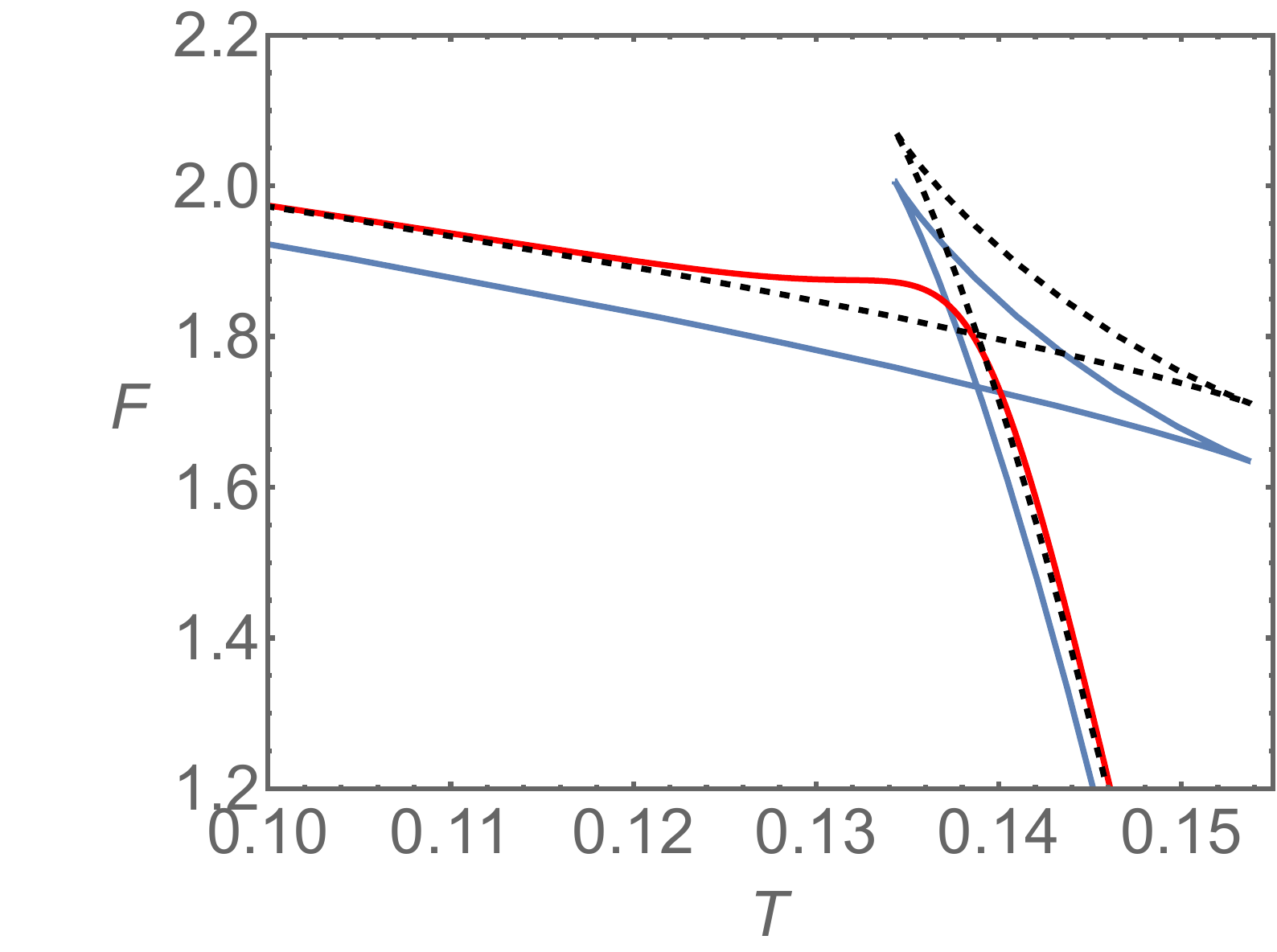}}
    \caption{Comparison of the quantum-corrected free energy $F_{\text{corrected}}$ and the ensemble-averaged free energy $\ave{F}$ with $G_{\N}=1/50$ and $G_{\N}= 1/10$, respectively.
    The red lines illustrate the ensemble-averaged free energy; the dashed black lines are the subleading-order quantum-corrected free energy, and the blue lines show the classical Gibbs free energy of the black hole. The angular momentum and pressure are set to be $J=0.05/G_{\N}$ and $P=0.5~P_c$.}
    \label{fig:one-loop-corre}
\end{figure}

For relatively small $G_{\N}$, the quantum-corrected free energy can be approximated as
\be
F_{\text{corrected}}=F_{\text{Gibbs}}+\frac{T_\h }{2}
\ee
In principle, we can regard the corrected free energy as the starting point of the whole thermodynamics. The corrected free energy can be further expressed as
\bea
F_{\text{corrected}}&=&\frac{1}{4G_{\N} \Xi r_\h}\left[a^2\Big(1-\frac{r_\h^2}{L^2}\Big)+\frac{2(a^2+r_\h^2)(L^2-r_\h^2)}{L^2-a^2}-r_\h^2\Big(1+3\frac{r_\h^2}{L^2}\Big)\right]\nonumber\\
&~&+\frac{r_\h}{8\pi\left(r_\h^2+a^2\right)} \left(1+\frac{a^2}{L^2}+\frac{3r_\h^2}{L^2}-\frac{a^2}{r_\h^2}\right)\,.\label{correctedF}
\eea
In the semi-classical limit with $G_{\N}\to 0$, the correction term in Eq.~\eqref{correctedF} can be ignored, and the classical black hole thermodynamics can be recovered.
As $G_{\N}$ increases, the ensemble-averaged free energy further deviates from the classical free energy, indicating that the quantum correction terms become more important. For large $G_{\N}$, the influence of the higher-order correction terms becomes more significant and the subleading correction alone is not sufficient to match the averaged free energy.
At this point, it becomes necessary to include additional sub-subleading terms.

In Fig. \ref{fig:one-loop-corre}, we show the first-order quantum-corrected free energy and the exact ensemble-averaged free energy with different Newton's constant $G_{\N}$.
As illustrated, the quantum-corrected free energy also exhibits swallowtail behaviors for $P=0.5P_c$ in the $F-T$ diagrams.
The quantum-corrected free energy aligns more closely with the ensemble-averaged free energy than the Gibbs free energy. The ensemble-averaged free energy should be regarded as the physical free energy when more quantum geometries are taken into account. In this context, the quantum-corrected thermodynamics offers a more accurate portrayal of the black hole system, incorporating additional quantum effects and exhibiting phase transitions.

We have an extra comment on the corrected thermodynamics. It is worth noting that the tree-level scattering amplitude is proportional to the vertices, and subleading-order terms of $G_{\N}$ capture the loop corrections in quantum gravity theory.
Thus, the finite $G_{\N}$ effects in the Euclidean path integral that we are trying to capture can be regarded as loop corrections to the black hole physics. In such a sense, we conducted an analytical study of the one-loop correction to black hole thermodynamics; however, higher-loop corrections require further investigation.

\section{Conclusion and discussion}
\label{sec:conclusion}

In this paper, we used the ensemble-averaged theory to study the thermodynamics of Kerr-AdS black holes.
In the conventional Kerr-AdS black hole thermodynamics, when the pressure of the AdS is smaller than the critical pressure, the Gibbs free energy exhibits a swallowtail behavior as it varies with the Hawking temperature.
We argued that this has a statistical interpretation from the ensemble-averaged theory.
By ensemble averaging over the probability distribution of states with conical singularity, we can derive the black hole phase transition in the thermodynamics limit. Effects beyond the thermodynamics limit can also be investigated.
This study offers a compelling framework for investigating the quantum properties of black holes and the nature of quantum gravity.

Inspired by thermal field theory, the density matrix of a theory can be obtained from the Euclidean path integral, which is also true for gravitational theory.
Based on the path integral approach, the generalized free energy for the Kerr-AdS black hole can be obtained, which incorporates the effects of all possible states with conical singularities in their Euclidean counterparts.
The extrema of the generalized free energy, which are saddle points of the Euclidean action, correspond to bulk black holes without conical singularity.
By introducing the density matrix inherited from the Euclidean gravitational path integral, we derived the probability distribution of different states. From this approach, we defined the ensemble-averaged free energy and explained how to sum over all contributions weighted by the probability distribution. We then argued that the quantum character of black hole thermodynamics is a finite $G_{\N}$ effect. By ensemble-averaging over all states, we showed that the black hole phase transition naturally arises in the small $G_{\N}$ limit.
It seems that the ensemble average is smart enough to pick the small/large black hole.
For large $G_{\N}$, the deviation from the classical saddles appears, which is regarded as a quantum effect from our perspective. We numerically calculated the ensemble-averaged free energy for different $G_{\N}$. Just as in the Schwarzschild-AdS and RN-AdS cases, away from the semi-classical regime, the ensemble-averaged theory exhibits a notable deviation from the conventional phase transition. We expanded the ensemble-averaged free energy in terms of $G_{\N}$ numerically, yielding the leading- and subleading-order contributions to the ensemble-averaged free energy. The leading-order ($O(G_{\N}^{-1})$) is just the Gibbs free energy of the black hole, from which the whole classical black hole thermodynamics can be derived. On the other hand, the sub-leading order ($O(G_{\N}^{0})$) contribution is half of the Hawking temperature, and can be regarded as a quantum correction. We can also approximate the probability distribution as a Gaussian around the peak and analytically calculate the leading and subleading terms, which are in agreement with the numerical results. This demonstrates that the Gaussian distribution is a valid approximation for the distribution function as long as $G_{\N}$ is small. Furthermore, we proved that the $T_\h /2$ correction is a universal feature inherent to black holes in asymptotically AdS spacetime.
When the subleading-order correction to classical thermodynamics is considered, the black hole thermodynamic should be modified accordingly.
We have shown that there are also swallowtail behaviors in the $F-T$ phase diagram for the quantum-corrected thermodynamics. Moreover, classical thermodynamics can be recovered in the semi-classical limit.



We believe the physics considered here is vital thus should be included in the black hole thermodynamics, due to the following reasons.
i) Those geometries provide a free energy landscape connecting different phases in black hole thermodynamics. With the semi-classical geometries as the saddles and the universal subleading-order behavior, the theory should be correct.
ii) The extended phase space with more geometries added in our study has a well-defined statistical interpretation, which is consistent with the gravitational path integral and matches our expectation of a statistical theory of black hole thermodynamics.
iii) The physics returns to the physics where we are familar with in the small $G_\text{N}$ limit.
So, it is natural to adjust $G_\text{N}$ to finite value and get what we have got in this paper.


It is worth noting that in the ensemble-averaged theory, we are working with the given ensemble temperature and the same Lorentzian physics. The allowed different bulk geometries are black holes with different sizes. Due to the given ensemble temperature, there are conical singularities for the geometries unless the Hawking temperature of the geometry equals the ensemble temperature.
In the thermodynamic phase space, we usually define the expectation value of a thermodynamic variable by integrating over a pair of conjugate canonical variables. It has been shown in the free energy landscape theory that the conjugate momentum should be $\dot{r}_{\text{h}}$, and therefore it is natural to use $\rh$ as the canonical variable in defining the ensemble average.
Moreover, $\rh$ is typically regarded as the order parameter and it characterizes different states of black holes. Using $\rh$ in the free energy landscape correctly captures the black hole phase transition.
In a sense, we have set an ambitious goal in \eqref{Zg} to integrate over all possible metrics in the path integral, yet have taken only a small step by extending the phase space with a single parameter $\rh$. This is a bottom-up logic in exploring the path integral formula.


It would be intriguing to extend the ensemble-averaged theory to modified gravity theories.
Note that in this paper, we use one single parameter to characterize the physics away from the saddle points of the action.
In general, we would like to explore more off-shell states in the gravitational path integral.
These so-called off-shell geometries are off-shell because the gravitational action is modified by adding terms that correspond to conical singularities.
Modified gravities offer more possible modifications to the action, meaning we may have more parameters to characterize the states away from the saddle points.
In this sense, we may get higher dimensional free energy landscapes rather than a simple one-parameterized generalized free energy.
The physics would also be much richer.
Given that the theory is grounded in the established principles of black hole thermodynamics and the Euclidean action, a well-defined Euclidean action for a modified gravity would allow us to employ the same methodology to extend the theory with relative ease.


In the semi-classical limit where we have finite $\hbar$ and small $G_N$, the physics is well described by quantum field theory (QFT) in a fixed spacetime.
By considering finite $\hbar$ effects, one can find logarithmic corrections to the black hole entropy in the QFT calculation.
However, here we focus on a different regime, where the corrections are from finite $G_N$ effects.
Rather than working with a fixed classical geometry, we aim to capture the effects of varying geometries in the path integral formulation in this work.

We have a final remark on the difference between the ensemble-averaged theory and the recently proposed dynamic evolution of the states from the free energy landscape~\cite{Li:2020khm,Li:2020nsy}. Although both of them describe the probability distribution of states, the evolution within the free energy landscape is predicated on the assumption that black hole phase transitions occur as a stochastic evolution process, as described by the Fokker-Planck equation. It is the stochastic evolution equation that governs the dynamics of the probability distribution. This framework encapsulates a non-equilibrium black hole thermodynamics, and the generalized free energy acts as an effective potential. 
However, in the ensemble-averaged theory, the density matrix is defined naturally through the path integral by incorporating conical singularities, and it describes the distribution of states. This approach differs significantly from the dynamic evolution of the probability density in the free energy landscape.



\begin{acknowledgments}
We would like to thank Bum-Hoon Lee, Yu-Xiao Liu, and Shao-Wen Wei for the helpful discussions.
This work is supported by the National Natural Science Foundation of China (Grant No. 12405073, No. 12305065, and No. 12247178), and the China Postdoctoral Science Foundation (Grant No. 2023M731468).
\end{acknowledgments}


\appendix

\section{A statistical interpretation of the thermodynamics via Euclidean path integral}
\label{boson}

For a bosonic system, we can write the Hamiltonian as a functional of the field $\phi(\textbf{x})$ and its conjugate momentum $\pi(\textbf{x})$:
\be
H=\int d^3 x~ \mathcal{H}[\pi(\textbf{x}),\phi(\textbf{x})]\,.
\ee
If we start with a state $\ket{\phi_a}$ at $t=0$, the state evolves into $e^{-i Ht}\ket{\phi_a}$ after a time interval $t$.
The transition amplitude for going from a state $\ket{\phi_a}$ to a state $\ket{\phi_b}$ after a time $t$ is
\be
\bra{\phi_b}e^{-it H}\ket{\phi_a}\,.\label{ta}
\ee
The transition amplitude admits a path integral representation
\begin{equation}
    \begin{split}
        \bra{\phi_b}e^{-it H}\ket{\phi_a}&=\int[\mathcal{D} \pi]\int_{\phi(\textbf{x},0)=\phi_a(\textbf{x})}^{\phi(\textbf{x},t)=\phi_b(\textbf{x})}[\mathcal{D} \phi]\\
&\times \exp\left[i\int_{0}^{t}d t\int d^3 x\left(\pi(\textbf{x},t) \frac{\partial\phi(\textbf{x},t)}{\partial t}-\mathcal{H}[\pi(\textbf{x},t),\phi(\textbf{x},t)]\right)\right]\,,
    \end{split}
\end{equation}
where we have a spacetime path integral.
For a theory with quadratic momentum term in the Hamiltonian, the transition amplitude can be further expressed in terms of the Lagrangian density $\mathcal{L}$
\bea
\bra{\phi_b}e^{-it H}\ket{\phi_a}=\int_{\phi(\textbf{x},0)=\phi_a(\textbf{x})}^{\phi(\textbf{x},t)=\phi_b(\textbf{x})}[\mathcal{D} \phi]~ \exp\left[i\int_{0}^{t}d t\int d^3 x ~\mathcal{L}\right]=\int^{\phi_b}_{\phi_a}[\mathcal{D} \phi]~ e^{iS}\,.
\eea

In equilibrium statistical mechanics, the situation is quite similar.
The canonical partition function can be expressed as
\be
Z_{\phi}=\tr~e^{-\beta H}=\sum_{\phi} \bra{\phi}e^{-\beta H}\ket{\phi}\,.\label{pf}
\ee
The summation in \eqref{pf} means summing over all possible $\phi$ configurations.
Elements of the statistical density matrix can be written as
\be
\bra{\phi_b}\rho \ket{\phi_a}=\bra{\phi_b}e^{-\beta H} \ket{\phi_a}\,.\label{dm}
\ee
Comparing the transition amplitude \eqref{ta} and the density matrix \eqref{dm}, it is not hard to find the similarity. The density matrix can be regarded as an Euclidean version of the transition amplitude with $\tau=-it$.
In the partition function calculation, we only care about the states that return to their original states after an Euclidean time interval $\beta$.
It is natural to associate a path integral description of the density matrix and partition function mimicking the transition amplitude formula. We have
\begin{equation}
    \begin{split}
        \bra{\phi_b}e^{-\beta H}\ket{\phi_a}=\int_{\phi(\textbf{x},0)=\phi_a(\textbf{x})}^{\phi(\textbf{x},\beta)=\phi_b(\textbf{x})}[\mathcal{D} \phi]~ \exp\left[-\int_{0}^{\beta}d \tau \int d^3 x ~\mathcal{L}\right]=\int_{\phi_a}^{\phi_b}[\mathcal{D} \phi]~ e^{-I_{\text{E}}}\,,\label{densitya}
    \end{split}
\end{equation}
with Euclidean action $I_{\text{E}}$.
The partition function can be expressed as
\bea
Z_{\phi}=\sum_{\phi}\bra{\phi}e^{-\beta H}\ket{\phi}=\int_{\text{periodic}}[\mathcal{D} \phi]~ e^{-I_{\text{E}}}\,,\label{Zphia}
\eea
where the word ``periodic'' means that we need to sum over all possible configurations that respect the periodic boundary condition in the $\tau$ direction.

\providecommand{\href}[2]{#2}\begingroup\raggedright\endgroup
\end{document}